\documentclass[traditabstract]{aa}
\usepackage{txfonts}
\usepackage{natbib}
\usepackage{graphicx,multirow}
\usepackage{color}
\newcommand{\Fe}{\mbox{\ion{Fe}{ii}}}
\newcommand{\Ox}{\mbox{[\ion{O}{iii}]}}
\newcommand{\Ni}{\mbox{[\ion{N}{ii}]}}
\newcommand{\Hb}{\mbox{H$\beta$}}
\newcommand{\Ha}{\mbox{H$\alpha$}}
\newcommand{\HII}{\mbox{\ion{H}{ii}}}

\newcommand{\text}{\mathrm}
\bibpunct{(}{)}{;}{a}{}{,}

\begin{document}
 
\titlerunning{The low-metallicity QSO HE~2158$-$0107}
\title{The low-metallicity QSO HE~2158$-$0107: A massive galaxy growing by the accretion of nearly pristine gas from its environment?
\thanks{Based on observations collected at the Centro Astron\'omico
Hispano Alem\'an (CAHA) at Calar Alto, operated jointly by the Max-
Planck-Institut f\"ur Astronomie and the Instituto de Astrof\'isica de
Andaluc\'ia (CSIC) and on observations collected at the European Organisation for Astronomical Reasearch in the Southern Hemisphere, Chile (program 070.B-0418).}}

\author{Bernd Husemann\inst{1}
   \and Lutz Wisotzki\inst{1}
   \and Knud Jahnke\inst{2}
   \and Sebastian F. S\'anchez\inst{3} 
}

\institute{Leibniz-Institut f\"ur Astrophysik Potsdam (AIP), An der Sternwarte 16, 14482 Potsdam, Germany\\
\email{bhusemann@aip.de}
\and 
Max-Planck-Institut f\"ur Astronomie, K\"onigsstuhl 17, D-69117 Heidelberg, Germany
\and
Centro Astron\'omico Hispano Alem\'an de Calar Alto (CSIC-MPIA), E-4004 Almer\'ia, Spain
}

\date{}
\abstract{The metallicities of Active Galactic Nuclei are usually  well above solar in their narrow-line regions, often reaching up to several times solar in their broad-line regions independent of redshift. Low-metallicity AGN are rare objects which have so far always been associated with low-mass galaxies hosting low-mass black holes ($M_\mathrm{BH}\lesssim 10^{6}\mathrm{M}_{\sun}$). In this paper we present integral field spectroscopy data of the low-redshift ($z=0.212$) QSO HE~2158$-$0107 for which we find strong evidence for sub-solar NLR metallicities associated with a massive black hole ($M_\mathrm{BH} \sim 3\times10^{8}\mathrm{M}_{\sun}$). The QSO is surrounded by a large extended emission-line region reaching out to 30\,kpc from the QSO in a tail-like geometry. We present optical and near-infrared images and investigate the properties of the host galaxy. The host of HE~2158$-$0107 is most likely a very compact bulge-dominated galaxies with a size of $r_\mathrm{e}\sim 1.4$\,kpc. The multi-colour SED of the host is rather blue, indicative of a significant young age stellar population formed within the last 1\,Gyr. A $3\sigma$ upper limit of $L_{\mathrm{bulge},H}<4.5\times10^{10}\mathrm{L}_{\sun,H}$ for the $H$ band luminosity and a corresponding stellar mass upper limit of $M_{\mathrm{bulge}}<3.4\times10^{10}\mathrm{M}_{\sun}$ show that the host is offset from the local black hole-bulge relations. This is independently supported by the kinematics of the gas. Although the stellar mass of the host galaxy is lower than expected, it cannot explain the exceptionally low metallicity of the gas. We suggest that the extended emission-line region and the galaxy growth are caused by the infall of nearly pristine gas from the environment of the QSO host. Minor mergers of low-metallicity dwarf galaxies or the theoretically predicted smooth accretion of cold ($\sim$\,$10^4$\,K) gas are both potential drivers behind that process. Since the metallicity of the gas in the QSO narrow-line region is much lower than expected, we suspect that the external gas has already reached the galaxy centre and may even contribute to the current feeding of the black hole.  HE~2158$-$0107 appears to represent a particular phase of substantial black hole and galaxy growth that can be observationally linked with the accretion of external material from its environment.}


\maketitle

\section{Introduction}
Active Galactic Nuclei (AGN) are thought to be powered by the accretion of ambient gas onto a supermassive Black Hole (BH) in the centre of their host galaxies. Because AGN can reach high luminosities for the most massive BHs ($M_\mathrm{BH}\sim10^{10}\mathrm{M}_{\sun}$), they can be observed as Quasi-Stellar Objects (QSOs) up to redshifts beyond $z>6$ \citep{Fan:2003,Mortlock:2011}. The emission lines in the Broad Line Regions (BLR) and the Narrow Line Regions (NLR) of AGN can be used to infer gas metallicities  \citep[e.g.][]{Hamann:1993}. QSOs are therefore considered as an important diagnostic for the metal enrichment and evolution of galaxies across cosmic time \citep[e.g.][]{Matteucci:1993,Hamann:1993,Hamann:1999}. The metallicities in the BLR were found to be supersolar, reaching $10Z_{\sun}$ in some cases, and to be correlated with the AGN luminosity, accretion rate, and BH mass \citep[e.g.][]{Hamann:1993,Shemmer:2004,Warner:2004,Nagao:2006,Matsuoka:2011}, with almost no redshift evolution. This is usually interpreted in terms of vigorous star formation building up the massive galaxies in which BHs reside and grow. 

Because the BLR is confined to the very circumnuclear region on scales of $\lesssim$1\,pc \citep[e.g.][]{Kaspi:2000}, it is not yet clear how the BLR metallicities are linked to the evolution of their host galaxies. The Narrow Line Region (NLR), on the other hand, is spatially extended on kpc scales \citep[e.g.][]{Bennert:2002,Schmitt:2003b,Bennert:2006b,Bennert:2006a,Husemann:2008} and is certainly more  representative of the host galaxy properties. Emission-line ratios in nearby AGN are readily observable and their values have been calibrated to metal abundances via photoionisation models \citep[e.g.][]{Storchi-Bergmann:1998,Groves:2004}. Systematic studies of obscured AGN by \citet{Groves:2006} and \citet{Barth:2008}, based on spectroscopic data of the Sloan Digital Sky Survey \citep[SDSS,][]{York:2000}, revealed that AGN with NLR metallicities around solar and below with respect to the solar abundance set of \citet{Asplund:2005} are extremely rare and only found in low-mass galaxies hosting BHs with $M_\mathrm{BH}\lesssim 10^{6}\mathrm{M}_{\sun}$. Since only a few percent of all luminous AGN are found in host galaxies with stellar masses less than $10^{10}\mathrm{M}_{\sun}$ \citep[e.g.][]{Kauffmann:2003}, the BH mass-bulge mass relation \citet{Haering:2004} combined with the mass-metallicity relation of galaxies \citep[e.g.][]{Tremonti:2004} thus naturally explains the rareness of low-metallicity AGN.

The QSO HE~2158$-$0107 at $z=0.212$ was discovered by the Hamburg-ESO survey \citep[HES,][]{Wisotzki:2000}. A radio flux of 1.6\,mJy at 1.4\,GHz was measured for this QSOs by the FIRST survey \citep{Becker:1995}. In relation to its optical brightness ($B=16.7$) the object is classified as a radio-quiet QSO. The SDSS targeted this QSO due to its X-ray and radio properties as well and catalogued it as \object{SDSS J220103.13-005300.2}.  It is also listed in the QSO catalogue of \citet{Veron-Cetty:2010}, but not in the catalogue of \citet{Schneider:2010}. The public SDSS spectrum of this particular QSO is of rather poor quality. In this paper we present evidence for HE~2158$-$0107 being an unusually low-metallicity QSO for its high BH mass, with a likewise unusually low-mass host galaxy. We combine our own integral field spectrophotometry with archival optical and infrared photometry to explore the properties of this particular QSO in more detail.

Throughout this paper we assume a cosmological model with $H_0=70\,\mathrm{km}\,\mathrm{s}^{-1}\,\mathrm{Mpc}^{-1}$, $\Omega_\mathrm{m}=0.3$, and $\Omega_\Lambda=0.7$. The adopted cosmology corresponds to a physical scale of $3.45\,\mathrm{kpc}\,\mathrm{arcsec}^{-1}$ at the redshift of HE~2158$-$0107 ($z=0.212$). We will always use the AB system \citep{Oke:1974} throughout the paper or explicitly refer to Vega magnitudes if required for the sake of comparison.

\section{Observations and data reduction}
\subsection{Integral field spectroscopy}
We observed HE~2158$-$0107 with the Potsdam Multi-Aperture Spectro-Photometer \citep[PMAS,][]{Roth:2005} mounted on the 3.5\,m telescope of the Calar Alto Observatory in the night of September 6th, 2002, as part of a larger sample of radio-quiet QSOs \citep[][Husemann et al. in prep.]{Husemann:2008}. Two exposures of 1800\,sec each were obtained under photometric conditions at a seeing of $1\farcs1$. 

In the lens array configuration of PMAS we covered an $8\arcsec\times8\arcsec$ field of view centred on the QSO. 256 individual spectra were obtained at a spatial sampling of $0\farcs5\times0\farcs5$. The low resolution V300 grism was positioned such that the \Hb\ and \Ha\ emission lines could be covered in a single exposure. We determined a spectral resolution for this instrumental setup of $R\sim900$ from a \ion{O}{i}\,$\lambda 5577$ night sky line width of 6.1\,\AA\ FWHM. 

The reduction of the integral field data was done with the \texttt{R3D} package \citep{Sanchez:2006a}. Here we just briefly summarise the process as the details will be described in detail in Husemann et al. (in prep.) where the entire dataset will be presented. We used the continuum and arc lamp calibration frames taken specifically for the target, at the same time and airmass, to trace and extract the 256 spectra on the CCD and to perform an absolute wavelength calibration. Twilight observations were obtained to construct a fibre flat-field to correct for the difference in fibre-to-fibre transmission. Photometric standard star observations allowed an absolute photometric calibration of the data including the correction for the atmospheric extinction at the Calar Alto site \citep{Sanchez:2007b}. A high S/N sky spectrum was extracted from a blank sky area \emph{within} the PMAS field of view, which we subsequently subtracted from the entire datacube. After correcting for differential atmospheric refraction we spatially aligned the two exposures and co-added both to obtain the final datacube.

\subsection{Near-Infrared and optical ground-based imaging}
An $H$ band image of HE~2158$-$0107 was acquired with the SOFI IR spectrograph and imaging camera \citep{Moorwood:1998} mounted on the New Technology Telescope of the La Silla Observatory. Ten exposures were taken in the large field camera mode (0\farcs29\,$\mathrm{pixel}^{-1}$) consisting of 12 sub-integrations (NDIT) with exposure times of 5\,sec (DIT) each, yielding a total integration time of 600\,sec on the source. The seeing during the observation was 0\farcs8. We reduced these images with standard calibration frames using the public SOFI reduction pipeline provided by ESO. We measured an $H$ band magnitude of 14.67\,mag (Vega) for the QSO which is quite close to the value of 14.70\,mag (Vega) reported by UKIDSS \citep{Warren:2007} and indicates a good photometric calibration of the image. 

Fortunately, HE~2158$-$0107 is in the footprint of the Supernova Survey as part of SDSS \citep[][for DR7]{Adelman-McCarthy:2008}, which covered 270 deg${}^2$ (a $2.5\degr$ wide scan along the celestial equator, generally called ``Stripe 82'') that were observed $\sim$80 times in the autumn of the years 1998--2007. We stacked the multiple epoch exposures in order to reach significantly deeper. This method was already applied to Stripe 82 data for precise photometry of galaxies \citep{Gilbank:2010} and to study details in the structure of nearby galaxies \citep{Schawinski:2010b}. Since observations of Stripe 82 were done under various ambient conditions, we co-added only the best 24 epochs with a seeing of $<$1\farcs3  and a limiting sky surface brightness of $>$20\,mag\,$\mathrm{arcsec}^{-2}$ in the $r$ band. The combined images effectively reach $\sim$5 times deeper ($\sim$2\,mag) compared to the individual SDSS images. Table~\ref{he2158_tab:sdss_image} summarises the effective seeing and limiting surface brightness for the combined images as well as the Galactic extinction inferred from the \citet{Schlegel:1998} extinction maps. 
\begin{table}
 \centering
 \caption{Properties of the SDSS Stripe 82 images}
  \label{he2158_tab:sdss_image}
 \begin{tabular}{cccc}\hline\hline\noalign{\smallskip}
Band  &  Seeing & Limiting mag &  $A$ \\ 
      &  [\arcsec] & [mag$\,\mathrm{arcsec}^{-2}$]  & [mag] \\ \noalign{\smallskip}\hline\noalign{\smallskip}
$u$ & 1.4 & 26.88 & 0.38 \\
$g$ & 1.2 & 27.54 & 0.28 \\
$r$ & 1.1 & 27.60 & 0.20 \\
$i$ & 1.0 & 26.94 & 0.15 \\
$z$ & 1.0 & 25.60 & 0.10 \\
\noalign{\smallskip}\hline\end{tabular}

\end{table}

\section{Black hole mass and NLR metallicity}
\subsection{ QSO spectrum of HE~2158$-$0107}
\begin{figure*}
  \includegraphics[width=\textwidth,clip]{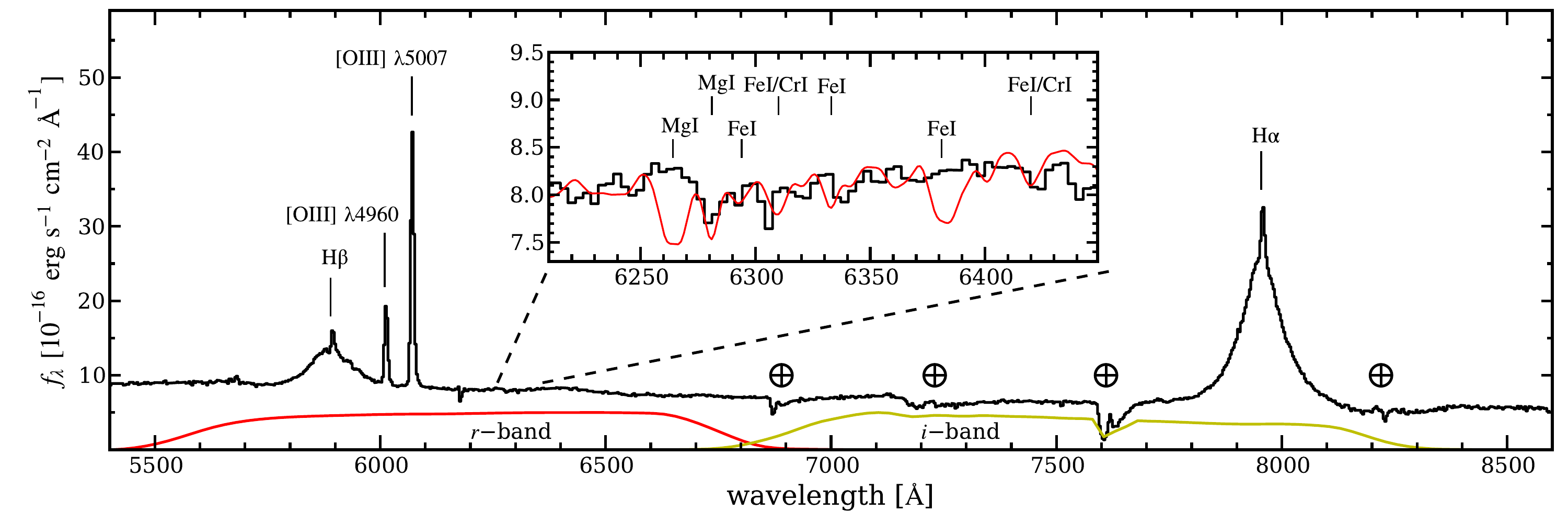}
  \caption{Integrated QSO spectrum of HE~2158$-$0107 as observed with the PMAS integral field spectrograph. All prominent emission lines are labelled and the four telluric absorption bands  in the red part of the spectrum are marked.  The red and yellow curves below the QSO spectrum represent the arbitrarily scaled $r$ and $i$ band transmission curves, respectively. The inset highlights the wavelength region of potentially strong stellar absorption lines. The redshifted spectrum of a 5\,Gyr old SSP with solar metallicity \citep{Bruzual:2003} is overplotted as the red line in the inset for direct comparison and the main chemical species leading to the various absorption lines are labelled.}
  \label{he2158_fig:qso_tot}
\end{figure*}

We first obtained the nuclear spectrum of HE~2158$-$0107 from the PMAS observation using a collapsed image of the datacube as a spatial weighting profile for the extraction. The resulting QSO spectrum with a S/N of about 100 per pixel in the continuum is shown in Fig.~\ref{he2158_fig:qso_tot}. Prominent broad \Hb\ and \Ha\ emission lines are visible as well as narrow emission lines from the NLR. From the narrow \Ox\ $\lambda5007$ line (\Ox\ hereafter) in the QSO spectrum we measured a redshift of $z=0.2118$. 
\begin{figure*}
\sidecaption
\includegraphics[width=12cm]{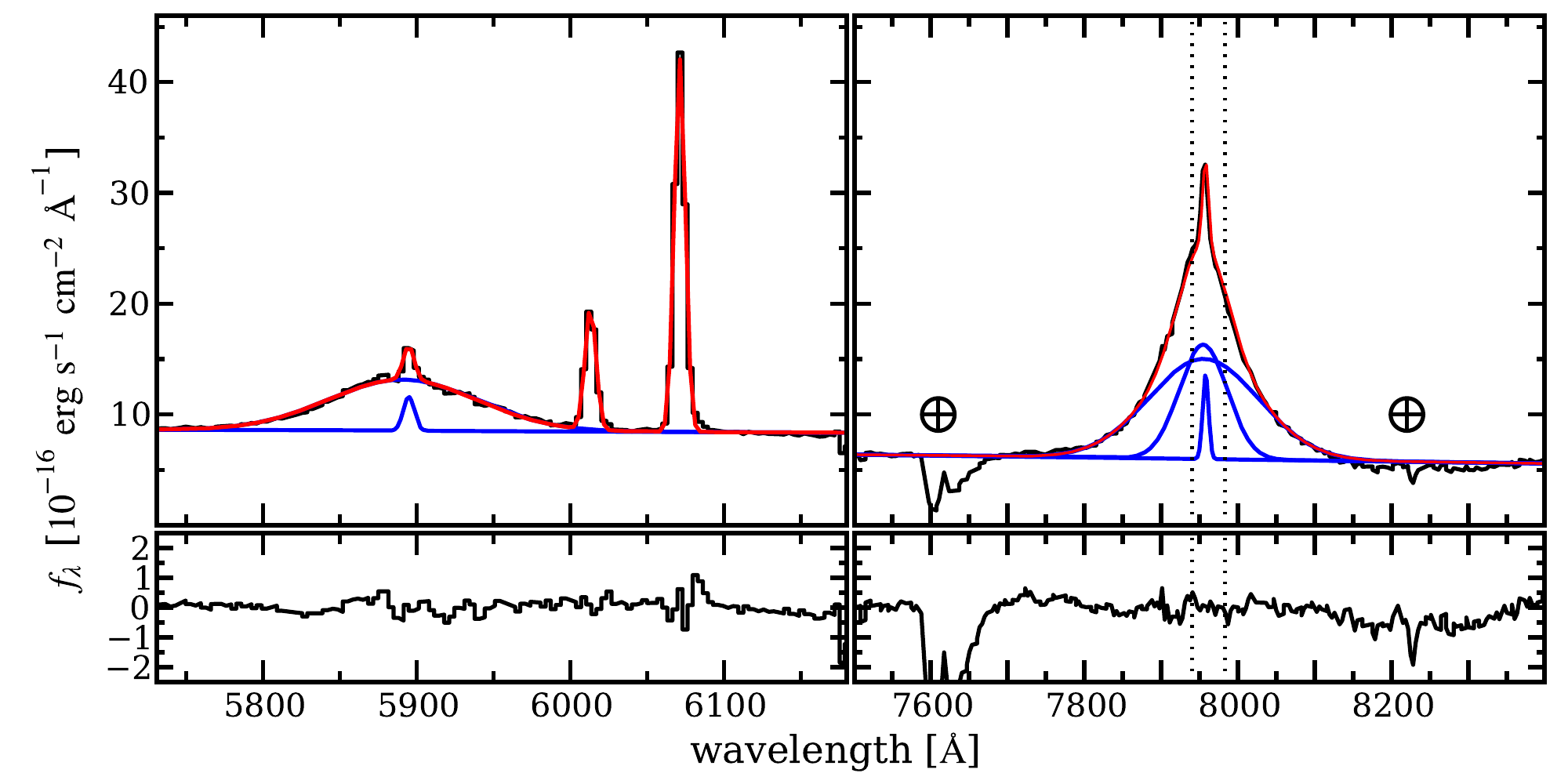}
 \caption{Modelling of the broad \Hb\ (left panel) and \Ha\ line (right panel). The red line represents the best-fit model to the data (black line) composed of individual Gaussians and a straight to approximate the local continuum. Individual Gaussian components for the Balmer lines are shown as blue lines above the continuum for a visual impression of their relative contribution. Two vertical dotted lines mark the wavelengths of the expected \Ni\,$\lambda\lambda6548,6583$ lines. The residuals of the best-fit model are displayed in the panels below. Some residuals of the bright \ion{Na}{i}$\,\lambda5893$ night sky line are visible on top of the broad \Hb\ line. Two telluric absorption bands of the Earth's atmosphere are clearly visible at 7600--7700\AA\ and 8150--8350\AA. }
 \label{he2158_fig:QSO_model}
\end{figure*}

 The synthetic $r$ band magnitude of the spectrum is brighter by only 0.13\,mag than the SDSS photometric $r$ band magnitude, which may well be caused by the intrinsic variability of the QSO. The \Fe\ complexes blue- and redward of \Hb\ \citep[e.g.][]{Boroson:1992} are very weak and do not contribute significantly to the QSO pseudo-continuum. We find tentative evidence for the presence of stellar \ion{Mg}{} and \ion{Fe}{}  absorption lines in the rest-frame wavelength  5120--5310\AA\ (see inset in Fig.~\ref{he2158_fig:qso_tot}). A properly redshifted 5\,Gyr old single stellar population (SSP) spectrum \citep{Bruzual:2003} matches with several spectral features in this wavelength range, although some stellar absorption lines seem to be absent. It is likely that those absorption lines are strongly blended in our low resolution spectrum with weak narrow emission lines of the QSO, like [\ion{Fe}{vii}] $\lambda5158$, [\ion{Fe}{vi}] $\lambda5176$, [\ion{N}{i}] $\lambda\lambda5177,5197$, or  [\ion{Fe}{vii}] $\lambda5278$. Thus, the nuclear QSO spectrum probably contain also a small fraction of stellar light from the underlying host galaxy.

\subsection{Virial BH mass estimate}
BH masses of QSOs can be estimated from their single epoch spectra via the so-called virial method \citep{Peterson:2000,Vestergaard:2002}. The method combines the empirically derived size-luminosity relation for the BLR, determined via reverberation mapping \citep[e.g.][]{Kaspi:2000,Peterson:2004,Kaspi:2005,Bentz:2006,Bentz:2009a},  and assuming virialised motion of the BLR clouds that can be inferred from the broad line width.

We used the formula also employed by \citet{Schulze:2010}, 
\begin{equation}
 M_\mathrm{BH}=2.57\times10^{7}\left(\frac{\sigma_{\mathrm{H}\beta}}{1000\,\mathrm{km}\,\mathrm{s}^{-1}}\right)^2\left(\frac{ L_{5100}}{10^{44}\mathrm{erg}\,\mathrm{s}^{-1}}\right)^{0.52}\,\mathrm{M}_{\sun}\ ,\label{he2158_eq:Hb_BHmass}
\end{equation}
where $\sigma_{\mathrm{H}\beta}$ is the line dispersion of the broad H$\beta$ line and $L_{5100}$ is the continuum luminosity at 5100\,\AA. Equation~(\ref{he2158_eq:Hb_BHmass}) is derived from the empirical relation between the BLR size and the continuum luminosity as calibrated by \citet{Bentz:2009a} combined with the prescription by \citet{Collin:2006} to infer the virial motions from $\sigma_{\mathrm{H}\beta}$ adopting a virial coefficient of $3.85$. By modelling the QSO spectrum with multiple Gaussians for the emission lines and a straight line for the local continuum (Fig.~\ref{he2158_fig:QSO_model}, left panel), we measured an \Hb\ line dispersion of $\sigma_{\mathrm{H}\beta}=2320\pm20\,\mathrm{km}\,\mathrm{s}^{-1}$, excluding any NLR component, and a continuum luminosity of $\lambda L_\lambda(5100)=(8.0\pm0.3)\times10^{44}\,\mathrm{erg}\,\mathrm{s}^{-1}$. From those measurements we estimated a BH mass of $\log(M_\mathrm{BH}/\mathrm{M}_{\sun})=8.6\pm0.3$ for HE~2158$-$0107.

\citet{Greene:2005} derived an alternative $M_\mathrm{BH}$ calibration based on the broad \Ha\ line luminosity and width, which is less sensitive to host galaxy light contamination and also less susceptible to dust extinction,
\begin{equation}
 M_\mathrm{BH}=3.0\times10^6\left(\frac{\mathrm{FWHM}_{\mathrm{H}\alpha}}{1000\,\mathrm{km\,s}^{-1}}\right)^{2.06}\left(\frac{L_{\mathrm{H}\alpha}}{10^{42}\,\mathrm{erg\,s}^{-1}}\right)^{0.45}\, \mathrm{M}_{\sun}\ .\label{eq:Ha_BHmass}
\end{equation}
We modelled the \Ha\ line with multiple Gaussian components (Fig.~\ref{he2158_fig:QSO_model}, right panel) and measured an \Ha\ luminosity of $(3.6\pm0.4)\times10^{43}\,\mathrm{erg\,s}^{-1}$ and a FWHM of $3860\pm70\,\mathrm{km\,s}^{-1}$ excluding the NLR narrow component.  This yields a BH mass of $\log(M_\mathrm{BH}/\mathrm{M}_{\sun})=8.4\pm0.3$ with Eq.~(\ref{eq:Ha_BHmass}). Although the broad \Hb\ line could be well approximated by a single Gaussian component, the \Ha\ model required two Gaussian components due to its extended wings that are not seen for \Hb. A difference in the line profiles is expected, because the FWHM of \Hb\ is systematically broader than \Ha\ \citep[e.g.][]{Greene:2005,Schulze:2010}. 

Both BH mass estimates differ only by 0.2\,dex and are well consistent with each other assuming a typical systematic uncertainty of $\sim$0.3\,dex for virial $M_\mathrm{BH}$ estimates. This shows that a high-mass BH powers the QSO nucleus of HE~2158$-$0107. We adopt the average of both estimate, $\log(M_\mathrm{BH}/\mathrm{M}_{\sun})=8.5\pm0.3$, for the rest of this paper.

\subsection{Evidence for an exceptionally low NLR gas-phase metallicity}
The estimated BH mass for HE~2158$-$0107 implies an expected bulge mass of $\log(M_\mathrm{bulge}/\mathrm{M}_{\sun})=11.3\pm0.3$ for its host galaxy adopting the empirical $M_\mathrm{BH}$-$M_\mathrm{bulge}$ relation of \citet{Haering:2004}. If the host galaxy has a substantial disc, the total mass of the system would even be higher, depending on the bulge-to-disc ratio. The mass-metallicity relation of galaxies implies that the gas-phase metallicity, usually expressed by the oxygen abundance, increases monotonically with stellar mass, so that we expect an oxygen abundance of $12+\log(\mathrm{O}/\mathrm{H})=9.13\pm0.2$ from the mass-metallicity relation of \citet{Tremonti:2004} for the inferred stellar mass of HE~2158$-$0107. However, different calibrations have been used in the literature to determine the oxygen abundance from strong emission lines of \ion{H}{ii} nebulae \citep[e.g.][]{Denicolo:2002,Kewley:2002,Pettini:2004,Kobulnicky:2004,Tremonti:2004,Pilyugin:2005}. Unfortunately, the absolute scale of the derived oxygen abundances depends strongly on the adopted calibration so that the corresponding mass-metallicity relations are also quite different \citep{Liang:2006,Kewley:2008,Calura:2009} even though the emission-line ratios are identical. A \emph{relative comparison} of abundances is therefore only possible when they were obtained with a common calibration.

For the radiation field of an AGN as in the case of HE~2158$-$0107, the oxygen abundance in the NLR region cannot be determined with the same calibrations used for \ion{H}{ii} regions. With state-of-the-art photoionisation codes like \texttt{CLOUDY} \citep{Ferland:1996}, emission line ratios can be interpreted by assuming a particular abundance pattern of elements, i.e. the solar abundance pattern, and a power-law ionising continuum of the AGN. Various studies reported that for the NLR the \Ox/\Hb\ ratio is primarily governed by the ionisation parameter while the \Ni/\Ha\ line ratio is mainly determined by the gas-phase metallicity \citep[e.g.][]{Storchi-Bergmann:1998,Groves:2004,Groves:2006}. A complication in that interpretation is that oxygen and nitrogen have a different nucleosynthesis. Oxygen is predominantly produced as a primary element in massive (young) stars based on the nucleosynthesis of hydrogen. Nitrogen can be produced similarly as a primary element, but it can also have a secondary origin in intermediate mass (old) stars generated from the original metals of the star \citep[e.g.][]{Matteucci:1986, Pilyugin:2003,Pettini:2008}. In this simple picture the ratio of N/O is constant if nitrogen has a primary origin, while it depends on O/H if it has a secondary origin. The secondary nature of nitrogen dominates at high metallicities $12+\log(\mathrm{O}/\mathrm{H})> 8.5$  \citep[e.g.][]{Vila-Costas:1993} where a relation between O/H and N/O can be empirically established.  At the centre of massive galaxies the metallicity of the gas is usually much higher than that limit. The linear relation 
\begin{equation}
 \log(\mathrm{N}/\mathrm{O}) = 0.96[12+\log(\mathrm{O}/\mathrm{H})]-9.29\label{eq:N_O}
\end{equation}
was obtained by \citet{Storchi-Bergmann:1994} from spectroscopic observations of galaxies with a nuclear starburst.

Assuming a secondary origin of nitrogen, \citet{Storchi-Bergmann:1998}  determined a calibration for the oxygen abundances of the NLR by matching the $\Ni\lambda\lambda\,6548,6583/\Ha\ (\equiv x)$ and $\Ox\lambda\lambda\,4960,5007/\Hb\ (\equiv y)$ line ratios with a grid of \texttt{CLOUDY} photoionisation models described by a two-dimensional polynomial of second-order,
\begin{eqnarray}
12+\log(\mathrm{O/H}) &=& 8.34 + 0.212x-0.012x^2-0.002y\nonumber\\
	&&+0.007xy-0.002x^2y+6.52\times10^{-4}y^2\nonumber\\
	&&+2.27\times10^{-4}xy^2+8.87\times10^{-5}x^2y^2\nonumber\\
	&&-0.1\log(n_\mathrm{e}/300)\ ,
\end{eqnarray}
where the dependence on the electron density ($n_\mathrm{e}$) was added as an additional term.

Since the oxygen abundance calibration used by \citet{Tremonti:2004} is also based on a match to \texttt{CLOUDY} photoionisation models, but for the conditions of \ion{H}{II} regions, the calibration of \citet{Storchi-Bergmann:1998} should provide oxygen abundances at roughly the same scale. We verify this by measuring the oxygen abundance of the NLR for all SDSS AGN having $\log10([\ion{O}{iii}]/\mathrm{H}\beta)>1.0$ and $\log10([\ion{N}{ii}]/\mathrm{H}\alpha)>-0.2$ using the method of \citeauthor{Storchi-Bergmann:1998}. We find a mean value of $12+\log(\mathrm{O}/\mathrm{H})=8.95\pm0.25$ which matches in absolute scale with the expected value for the mass-metallicity relations at a stellar mass of $\log(M_*)\approx10.5$ where the mass distribution of AGN host galaxy peaks \citep{Kauffmann:2003} for the \citet{Tremonti:2004}, \citet{Kobulnicky:2004}, and  \citet{Kewley:2002} calibrations whereas the other calibrations are substantially offset. We are therefore confident in using the mass-metallicity relation of \citeauthor{Tremonti:2004} as our reference for comparison with the NLR metallicity that we  determine in the following.
\begin{figure}
\resizebox{\hsize}{!}{\includegraphics[clip]{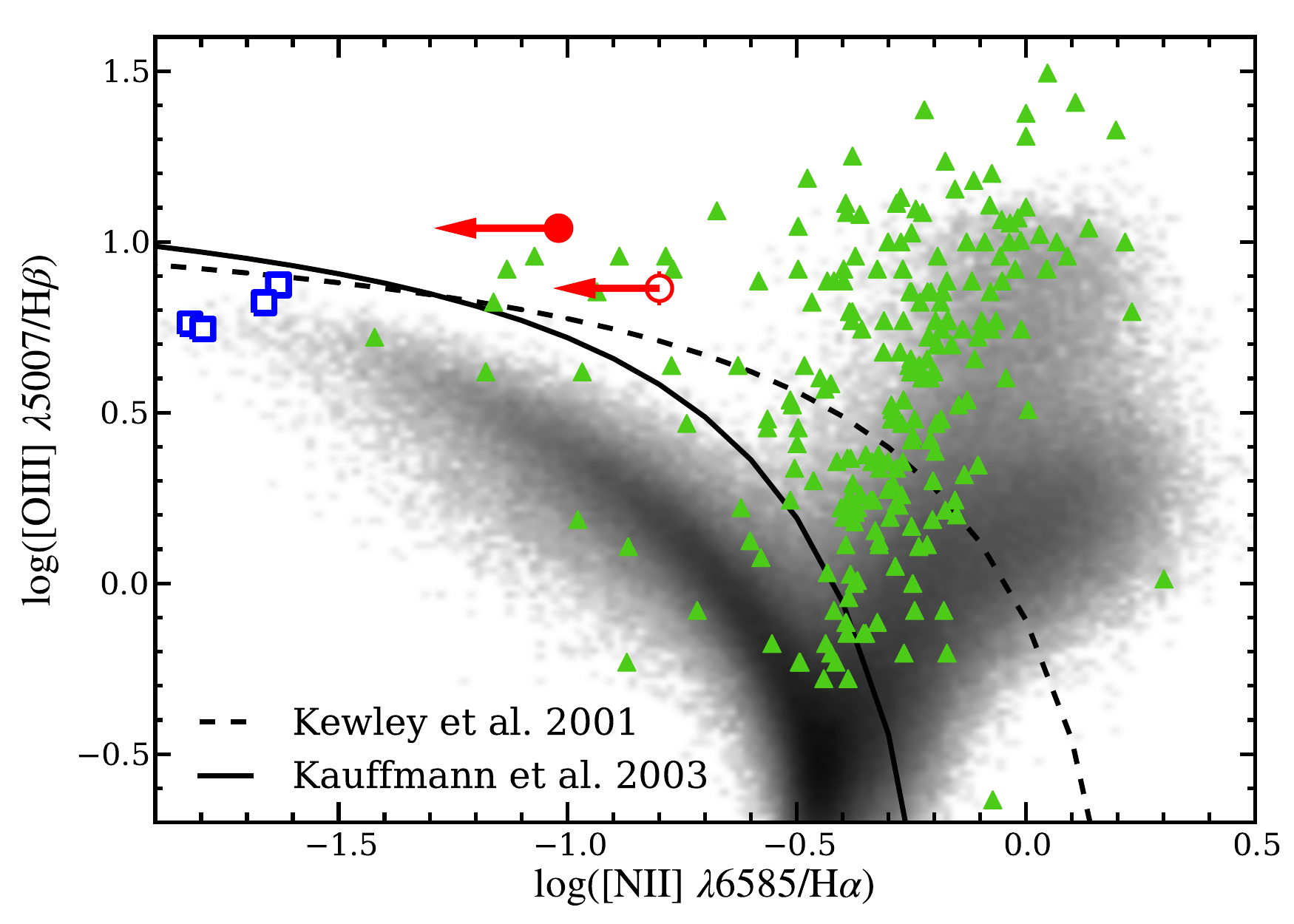}}
 \caption{Standard BPT emission-line ratio diagram for HE~2158$-$0107. The measured line ratios of the NLR (filled red circle) and the EELR (opened  red circle) are shown for the 3\,$\sigma$ limits on $\log(\Ni/\Ha)$ and the length of the arrows are such that their tip corresponds to the 1\,$\sigma$ limit. The distribution of emission-line ratios for $\sim$40000 randomly selected SDSS galaxies are indicated by the grey scale for comparison. Triangle symbols indicate the NLR line ratio for a sample of AGN with low BH masses \citep{Greene:2007} and blue opened squares represent four candidate low-metallicity AGN in dwarf galaxies as reported by \citet{Izotov:2008} and \citet{Izotov:2010}.}
 \label{he2158_fig:BPT_diagram}
\end{figure}

The QSO spectrum of HE~2158$-$0107 displays prominent narrow emission lines from the NLR and we included also Gaussian components for the narrow \Ni\,$\lambda\lambda 6548,6583$ emission line doublet to the model of the \Ha\ line (see Fig.~\ref{he2158_fig:QSO_model}). However, no trace of the \Ni\,$\lambda6583$ (\Ni\ hereafter) line is visible in the data. It is of course possible that weak \Ni\ line is actually present and just below our detection limit considering the low spectral resolution and the blending with the underlying broad \Ha\ line. In order to determine a realistic detection limit for the \Ni\ line, we performed Monte-Carlo simulation as described in the Appendix. From those simulations we determine a $3\sigma$ upper limit of $\log(\Ni/\Ha)<-1.0$, or  $\log(\Ni/\Ha)<-1.2$ at 1\,$\sigma$. The location of the NLR of HE~2158$-$0107 in the $\log(\Ox/\Hb)$ vs. $\log(\Ni/\Ha)$ diagram, the so-called BPT diagnostic diagram \citep{Baldwin:1981}, is shown in Fig.~\ref{he2158_fig:BPT_diagram}. The line ratios place the NLR in the domain of AGN ionisation above the \citet{Kewley:2001} and \citet{Kauffmann:2003} demarcation curves as expected, but at a significant offset from the local population of AGN towards a much lower \Ni/\Ha\ line ratio. With the abundance calibrations of \citet{Storchi-Bergmann:1998} we infer an upper limit of $12+\log(\mathrm{O}/\mathrm{H}) < 8.4$ for the oxygen abundance in the NLR of HE~2158$-$0107 based on the \Ox/\Hb\ and \Ni/\Ha\ line ratios and assuming a canonical electron density for the NLR of $n_\mathrm{e}\sim1000\,\mathrm{cm}^{-3}$. This oxygen abundance corresponds to $Z<0.5Z_{\sun}$ and also agrees with the gas-phase metallicity of the \citet{Groves:2006} photoionisation models.

\begin{figure*}
  \includegraphics[width=\textwidth,clip]{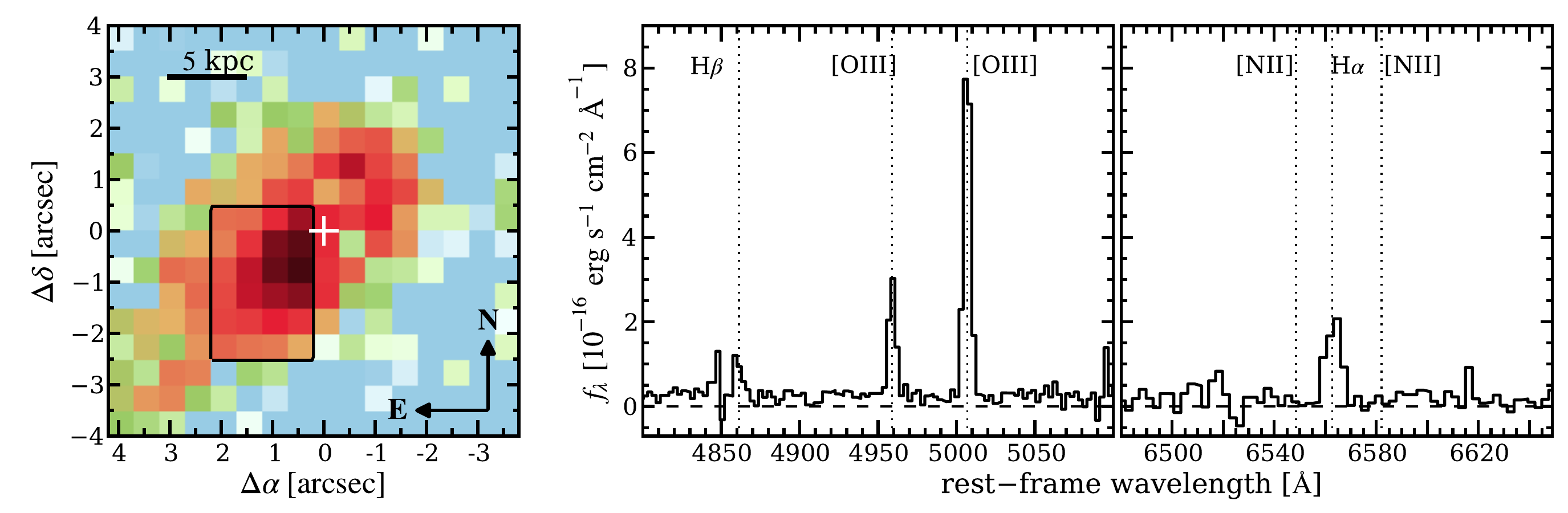}
  \caption{\Ox\ light distribution and EELR spectrum for HE~2158$-$0107. \textit{Left panel:} Nucleus-subtracted 40\AA\ wide \Ox\ narrow-band image as extracted from our PMAS datacube. The position of the QSO is highlighted by the white cross. A black box indicates the boundary for the extraction of a co-added spectrum. \textit{Right panel:} Co-added spectrum of the EELR region as defined in the left panel. The expected wavelengths of various emission-lines are indicated by the vertical dotted lines.}
  \label{he2158_fig:EELR_prop}
\end{figure*}

Our judgement that the exceptionally low \Ni/\Ha\ line ratio is not caused by the blending with the broad \Ha\ line is further supported by line ratios of an Extended Emission Line Region (EELR) that we resolved around the QSO (Fig.~\ref{he2158_fig:EELR_prop}). We employed our software tool \texttt{QDeblend${}^\mathrm{3D}$} for the deblending of the QSO and extended emission using an iterative algorithm improving the technique introduced by \citet{Christensen:2006}. Details on the algorithm can be found in the \texttt{QDeblend${}^\mathrm{3D}$} manual\footnote{http://sourceforge.net/projects/qdeblend/}, and its application to the PMAS IFU observation of our entire QSO sample will be described  more detailed in Husemann et al. (in prep.). Briefly, a high S/N QSO spectrum from the central spaxel was subtracted from all other spaxels after being scaled to match the broad Balmer lines. Three iterations were performed to empirically decontaminate the QSO spectrum from its host contribution using a co-added spectrum of the 8 spaxels surrounding the central QSO spaxel. We extracted a co-added spectrum from a high surface brightness region South-East of the QSO as indicated in Fig.~\ref{he2158_fig:EELR_prop}. No broad emission lines are left in the extended spectrum after we applied our QSO-host deblending process, so that the blending of broad and narrow lines is not a problem for the EELR spectrum. The \Ni\ doublet is still undetected in this spectrum and we determined a $3\sigma$ upper limit of \Ni/\Ha$<-0.8$ (\Ni/\Ha$<-0.94$ at 1$\sigma$) using the same type of simulation method applied for the QSO spectrum. The emission-line ratios of the EELR are close to the unresolved NLR in the BPT diagram (Fig.~\ref{he2158_fig:BPT_diagram}). It is likely that the AGN is the main ionisation source for the extended emission as well. However, due to the 0.15\,dex lower  \Ox/\Hb\ line ratio and the upper limit in \Ni/\Ha, we cannot strictly rule out excitation by young massive stars from emission-line diagnostics alone.  In either case, the limit on the \Ni/\Ha\ line ratio implies that the gas-phase metallicity must be low over a region of several kpc, suggesting that the low metallicity gas is not confined to the circumnuclear region. 

Only very few other AGN with similarly low oxygen abundances in the NLR gas are known to date. \citet{Groves:2006} found only 40 low-metallicity AGN among 23\,000 Seyfert 2 galaxies, all of which are in low-mass galaxies. Only two of them actually have gas-phase metallicities lower than the upper limits of HE~2158$-$0107. \citet{Greene:2007} constructed a sample of low BH mass AGN ($M_\mathrm{BH}<2\times10^6\mathrm{M}_{\sun}$) and measured the line ratios of the NLR which are shown in Fig.~\ref{he2158_fig:BPT_diagram} for comparison. They also found only 2 AGN with $\log(\Ha/\Ni)<-1.0$ in their NLR. Finally, 4 low-metallicity AGN candidates with $\log(\Ha/\Ni)\sim-1.6$ were reported by \citet{Izotov:2008} and \citet{Izotov:2010}, all residing in dwarf galaxies with BH masses in the range of the \citeauthor{Greene:2007} sample. In contrast to the other AGN considered so far, those candidate AGN display emission-line ratios in the NLR that are well consistent with classical \HII\ regions powered by massive stars. Therefore, they should be considered as a completely different class of objects.

HE~2158$-$0107 differs from the above mentioned cases in that it is a luminous QSO with a high-mass BH. It is therefore the first and so far only known proper case of a low-metallicity QSO. Our limiting gas-phase oxygen abundance is 0.7\,dex lower than expected for the stellar mass predicted by the BH mass-bulge mass relation. Either the host of  HE~2158$-$0107 is an extreme outlier from the mass-metallicity relation of galaxies, or it is strongly offset from the BH mass-bulge mass relation, or even both. We explore the latter aspect in the next section.

\section{The multi-colour SED of HE~2158$-$0107 and its bulge mass}
In order to directly constrain the bulge mass and to test whether the host galaxy of HE~2158$-$0107 is significantly offset from the 
BH mass-bulge mass relation, we analysed the optical and infrared broad-band images to reconstruct the multi-colour Spectral Energy Distribution (SED) of the host galaxy. To do this, the emission of the QSO and the its host need to be decomposed, which requires a good characterisation of the Point Spread Function (PSF) in each band. 

Two bright stars are close to HE~2158$-$0107, brighter than the QSO by 2\,mag and 1\,mag in the $r$ band, respectively. We empirically determined the PSF for each observation from these two stars. Their distances to the QSO of 62\arcsec\ and 76\arcsec\  already caused subtle PSF variations with respect to the PSF at the position of the QSO. These variations are particularly strong for the SOFI image taken with the large field objective due to coma \citep{Moorwood:1998}. To overcome the limitations imposed by the field variations we analysed their PSF residuals from another star only 18\arcsec\ away from the QSO, which has a similar  $r$ band magnitude. We used the residuals within a radius of 2\arcsec\ to make an empirical corrections to our two PSF stars. Since all the stars in the SDSS field are at least 0.5\,mag fainter than the QSO in the $u$ band, we were not able to construct a useful PSF in this band and excluded it from our analysis.

\subsection{Results of the QSO-host decomposition}
We employed \texttt{GALFIT v3.0.2} \citep{Peng:2002,Peng:2010} to decompose the host galaxy and the point-like QSO of HE~2158$-$0107 by modelling the observed two-dimensional surface brightness (SB) distribution in the SDSS and SOFI broad-band images with an appropriate model. We considered two different models. Model 1 consists of a single PSF, only to check whether any detectable signal from the host galaxy remains in the residuals even when the QSO contribution is over-subtracted. Model 2 is composed of a de Vaucouleur profile for the presumably bulge-dominated host and a PSF component for the QSO. After trying other models as well, the host turned out to be so faint that we fixed the SB model of the host to the de Vaucouleur profile to increase the robustness of the \texttt{GALFIT} model and the recovered host galaxy magnitudes \citep[e.g.][]{Sanchez:2004b,Kim:2008b}. Although other SB models for the host galaxy including a disc component might be possible, we emphasise that those models would inevitably lead to an even \emph{fainter} bulge component. Hence, our approach provides a robust upper limit for the bulge luminosity.
\begin{figure*}
 \includegraphics[width=\textwidth,height=19.0cm]{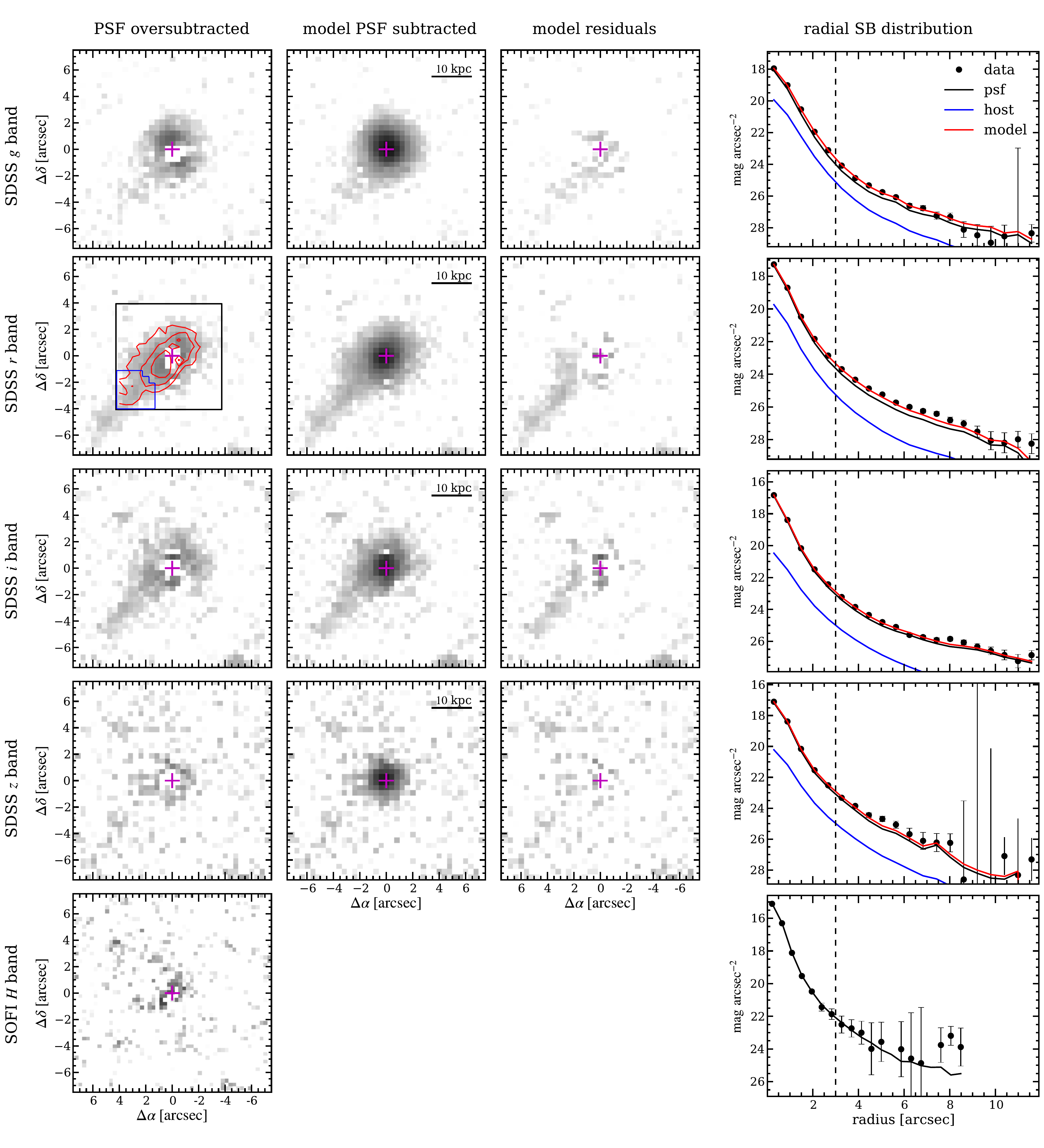}
 \caption{Host galaxy images of HE~2158$-$0107 after 2D QSO-host decomposition with \texttt{GALFIT}. The first column represents the PSF (over)subtracted host images of the best-fit model including a point source only. The PSF subtracted host images and residuals of the best-fit model, including a point source \emph{and} a de Vaucouleurs profile for the host, are shown in the second and third column, respectively. The radial surface brightness profiles of the data, best-fit model, and their components are presented in the fourth column to highlight the low contrast between nucleus and host. The orientation and position of the PMAS field of view with respect to the QSO is indicated by the black rectangle in the $r$ band image and the corresponding PMAS \Ox\ narrow-band image by the red contours. A region covering the extended tail analysed in Sect.~\ref{sect:tail} is marked by the blue boundary.}
 \label{fig:host_models}
\end{figure*}

\begin{table*}
 \centering
 \caption{QSO-host decomposition results corrected for systematics}
  \label{tab:host_models}
 \begin{tabular}{lccccccc}\hline\hline\noalign{\smallskip}
Passband	& $m_\mathrm{QSO}$ & $m_\mathrm{host}$ & $r_\mathrm{e}$     & $b/a$ & PA & $m_\mathrm{host}$ ($<3\arcsec$)\\
                &  [mag]           & [mag]             & [\arcsec] &   &[\degr]& [mag]    \\\hline\noalign{\smallskip}                                                   
SDSS $r$        & $16.18\pm0.01$   & $18.53\pm0.04$    & $0.52\pm0.03$ & 0.59  & 47 & $18.65\pm0.04$ \\
SDSS $i$        & $15.79\pm0.01$   & $18.71\pm0.05$    & $0.58\pm0.06$ & 0.63  & 40 & $18.91\pm0.05$ \\
SDSS $z$        & $15.99\pm0.02$   & $18.78\pm0.11$    & $0.40\pm0.10$ & 0.90  & 24 & $18.89\pm0.11$   \\\noalign{\smallskip}\hline
\end{tabular}

\end{table*}

In Fig.~\ref{fig:host_models} we present the results of the QSO-host decomposition. In all cases, except the $H$ band, we resolve the underlying host galaxy and recover the structural properties for our assumed host model. Interestingly, the $r$ and $i$ band host images reveal an extended tail in the light distribution, towards the South-East, that matches with the EELR seen in our PMAS data. We find that the ellipticity ($e\equiv1-b/a$) and effective radius $r_\mathrm{e}$ of the best-fit de Vaucouleur profile are larger in the $r$ and $i$ bands than in the $g$ and $z$ bands. 
This is expected if line emission of the EELR significantly contributes to the signal in these particular bands  (see Fig.~\ref{he2158_fig:qso_tot}).  In order to estimate the emission-line contribution of the  EELR to the $r$ and $i$ bands, we synthesised a pure emission-line spectrum for the EELR based on its \Ox\ flux of $(150\pm30)\times10^{-16}\,\mathrm{erg}\,\mathrm{s}^{-1}\,\mathrm{cm}^{-2}$ within an aperture of 3\arcsec\ centred on the QSO and the observed line ratios. From the synthetic EELR spectrum we determined an emission-line brightness in the $r$ and $i$ bands of $20.5\pm0.2$\,mag and $21.6\pm0.2$\,mag, respectively. This corresponds to a substantial emission-line contribution of $18\pm7\%$ within the central $3\arcsec$ of the $r$ band and a rest-frame \Ox\ equivalent width (EW) of $\mathrm{EW(\Ox)}=170\pm70$\,\AA.

Due to the apparently large nucleus-to-host ratio of $>$10 in the $z$-band and the small apparent size of the host galaxy, it is furthermore import to estimate the systematic errors and to clean the structural parameters and magnitudes from systematic effects. We performed a suite of Monte-Carlo simulations for various nucleus-to-host ratios and five different effective radii for each band.   We used one empirical PSF to generate 200 mock images for a given set of parameters and analysed the image subsequently with the other empirical PSF. Detailed information about these simulations are given in the Appendix. 

The \texttt{GALFIT} results of the best-fit models are listed in Table~\ref{tab:host_models} after correcting them for Galactic extinction and for the systematic biases recovered by the simulation. We determined a $3\sigma$ upper limit for the host magnitude of $m_\mathrm{host}>18.19$\,mag in the $H$ band from our simulation. We withdraw the $g$ band from this analysis because the QSO-host decomposition appears to be significantly affected by an intrinsic PSF mismatch between the stars and the QSO in this band. The SEDs of the QSO and the stars are so different in the blue that their ``effective'' PSFs are heavily affected by the wavelength dependence of the seeing. To quantify this effect we simulated synthetic $ugriz$ images. For simplicity we assumed a symmetric Gaussian profile for the PSF with a FWHM that is a smooth function of wavelength as constrained by the values in Table~\ref{he2158_tab:sdss_image}. Assuming a power-law spectrum for the QSO and a matched  stellar spectrum for the stars, we constructed monochromatic images for each wavelength (stepsize 1\AA). Finally, the band images were created by summing up the monochromatic images with its contribution set by the transmission curve of the band. Afterwards we scaled the stellar image to match with the QSO image in the brightest pixel and subtracted it. We find a residual flux of the order of 2\% for the $u$ and $g$ bands, but only $<0.1\%$ in the $r$, $i$, and $z$ bands. Since the nucleus-to-host ratio for HE~2158$-$0107 is already 10 in the $r$ band, such a residual flux solely due to PSF mismatch is quite significant.

\subsection{An SED-based stellar mass estimate}
\begin{figure}
 \resizebox{\hsize}{!}{\includegraphics[clip]{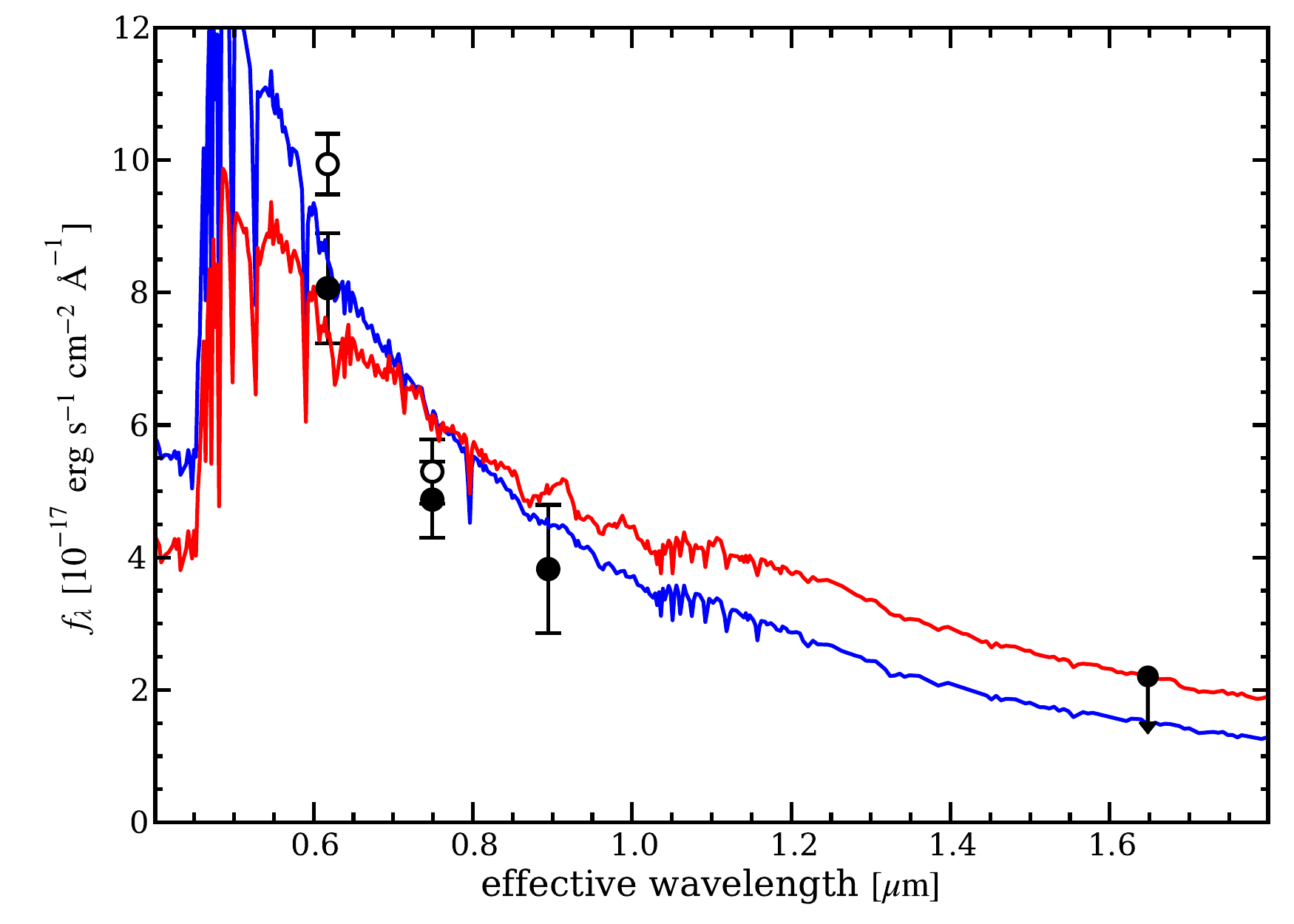}}
 \caption{The host galaxy SED of  HE~2158$-$0107 within a circular aperture of $3\arcsec$. Open and filled symbols correspond to the SED with and without the contribution of the emission lines in the $r$ and $i$ band. The best-fit 100\,Myr SSP model (blue line) and composite 100\,Myr+11\,Gyr SSP model (red line) taken from the \citet{Bruzual:2003} library of spectra are shown for comparison.}
 \label{fig:host_SED}
\end{figure}

After taking into account the systematics of the two-dimensional QSO-host decomposition, the emission-line contribution, and the foreground Galactic extinction in the different bands, we show the SED of the host galaxy from the $rizH$ photometry in Fig.~\ref{fig:host_SED}. Even after subtracting the EELR contribution from the SED, the shape of the SED appears to be very blue.  Scattered light of the QSO due to large amounts of dust within the host could significantly contribute to the apparent host light in rest-frame blue and UV bands \mbox{\citep[e.g.][]{Tadhunter:1992,Zakamska:2006}.} We thus looked at the \emph{GALEX} UV images of the Medium Imaging Survey for HE~2158$-$0107. They are consistent with a point source without any asymmetry that would indicate large scale scattered emission. Since the low gas-phase metallicity suggests a low dust content anyway, we think that a young stellar population as often found in bulge-dominated QSO hosts \citep[e.g.][]{Kauffmann:2003,Jahnke:2004b,Sanchez:2004b} accounts for the blue colour. 

In order to constrain the stellar population, we modelled the SED with a grid of SSP model spectra from \citet{Bruzual:2003} with ages in the range of 5\,Myr to 17\,Gyr and  $0.2Z_{\sun},\ 0.4Z_{\sun},\ 1Z_{\sun}$ and $2Z_{\sun}$ metallicities. A $\chi^2$ test was applied to check which SSPs models were consistent with the photometry at a 95\% confidence level. We incorporated the upper limit in the $H$ band by assuming a value of half the limit and a $3\sigma$ error matching with the $3\sigma$ limit. Given the few photometric data points and their considerable errors we find that formally all SSPs with ages younger than  1\,Gyr are consistent with the data. The best-fit SSP of solar metallicity has an age of 100\,Myr. From all matching SSP models we derive a minimum stellar mass of $M_\mathrm{bulge}>3.0\times10^{8}\mathrm{M}_{\sun}$. This extremely low value of course just constrains the young stellar population and is therefore useless for our purposes.

Although the light of the host galaxy is apparently dominated by young stars, the mass of the system could still be dominated by an underlying old stellar population. Thus, we determined how much stellar mass in an old 11\,Gyr SSP can be \emph{added} to the stellar mass of the young SSP, so that the composite spectrum still obeys the $3\sigma$ upper limit of $L_{\mathrm{bulge},H} < 4.5\times10^{10}\mathrm{L}_{{\sun},H}$ in the $H$ band. From all possible combinations we infer a maximum total stellar mass for the host galaxy of $M_\mathrm{bulge}<3.4\times10^{10}\mathrm{M}_{\sun}$.

\begin{figure}
 \resizebox{\hsize}{!}{\includegraphics[clip]{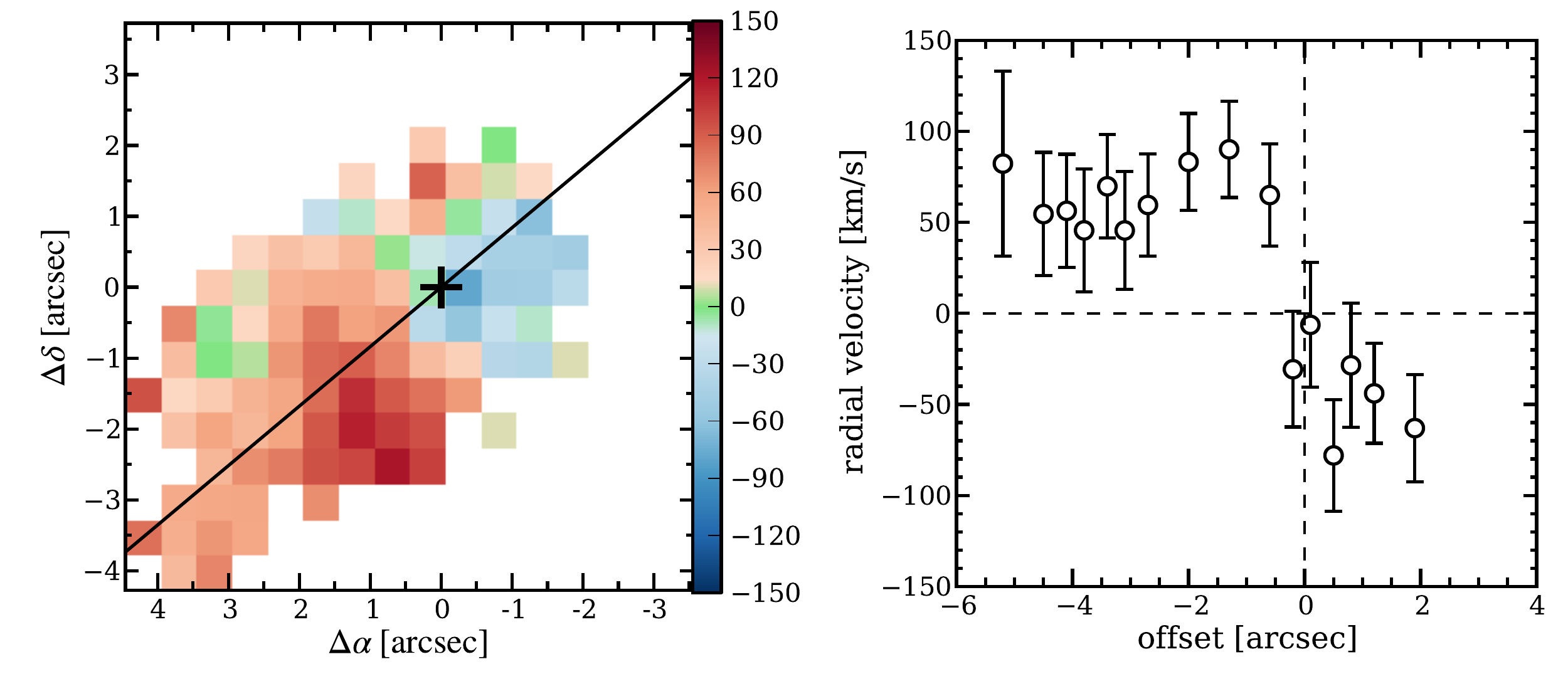}}
 \caption{Velocity field of the \Ox\ EELR around HE~2158$-$0107. \textit{Left panel:} Radial velocity map inferred from the \Ox\ emission line Doppler shift with respect to the rest-frame of the QSO. The black solid line represent roughly the apparent kinematic major axis. \textit{Right panel:} Radial velocity curves extracted from the velocity map along the kinematic major axis (black open circles).}
 \label{fig:EELR_vel}
\end{figure}

\subsection{Additional mass constraints from the gas kinematics}
In Fig.~\ref{fig:EELR_vel} we present the \Ox\ velocity field of the EELR which we derived by modelling the \Ox\ doublet lines in each spaxel of the nucleus-subtracted datacube with Gaussian profiles. While quite noisy, the overall appearance of the velocity field is consistent with ordered rotation. We constructed a synthetic longslit curve from the velocity map along the apparent kinematic major axis from all spaxels within $\pm0\farcs5$ away from it. The radial Doppler motion of the gas was estimated using the redshift of the spatially unresolved \Ox\ line as the reference for the QSO rest-frame. From the velocity curve along the major axis we infer a maximum radial velocity of $v_r=80\pm20\,\mathrm{km\,s}^{-1}$ at a radius of 2\arcsec\ from the nucleus. 

\begin{figure*}
 \includegraphics[width=\textwidth,clip]{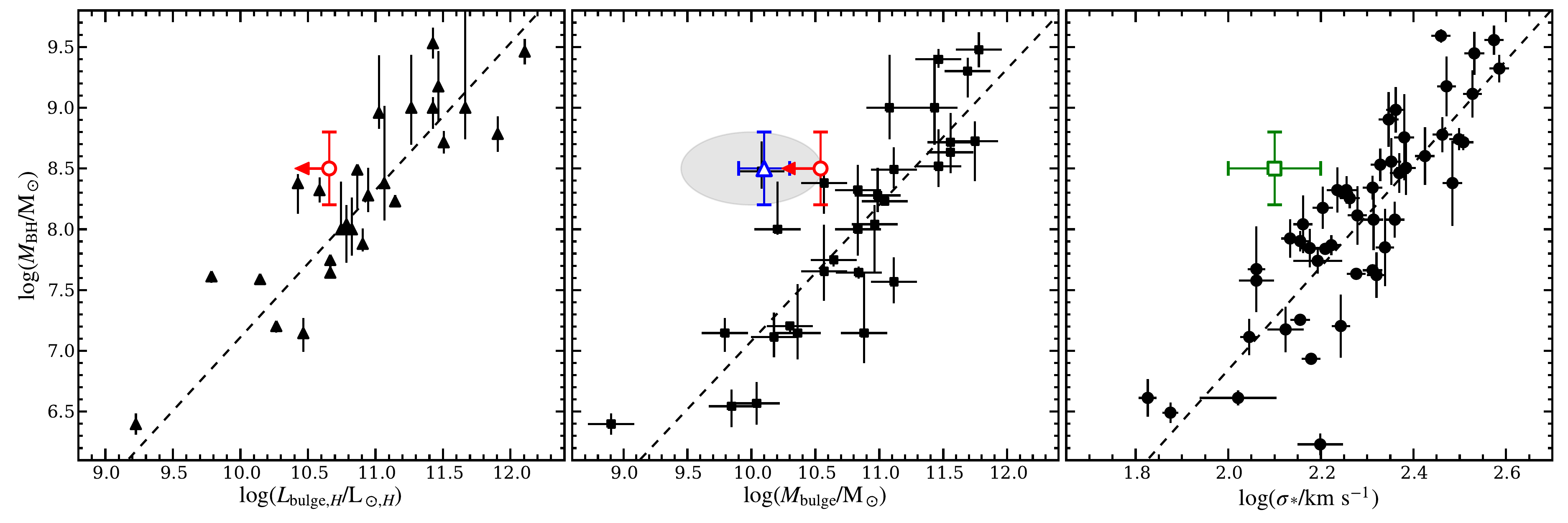}
 \caption{\emph{Left panel:} $M_\mathrm{BH}$-$L_\mathrm{bulge}$ relation for the $H$ band. The black triangles are measurements from \citet{Marconi:2003} for local galaxies and the dashed line is their best-fit relation. Our $3\sigma$ upper limit on $L_{H,\mathrm{bulge}}$ for HE~2158$-$0107 is shown as the red open circle. \emph{Middle panel:} $M_\mathrm{BH}$-$M_\mathrm{bulge}$ mass relation. The black squares correspond to the measurements of \citet{Haering:2004} with their best-fit relation shown as the dashed line. An SED based $3\sigma$ upper limit for the stellar mass of the HE~2158$-$0107 is indicated by the red open circle. An independent estimate based on the bulge size in the $z$-band using the size-mass relation by \citet{Shen:2003} is shown as the blue open triangle. The light grey area represents the possible range in $M_\mathrm{dyn}$ of the bulge based on the ionised gas kinematics assuming rotational motion of a cold gas disc. \emph{Right panel:}  $M_\mathrm{BH}$-$\sigma_\mathrm{*}$ relation for local galaxies with dynamical $M_\mathrm{BH}$ measurements. Measurements by \citet{Gueltekin:2009} are shown as black round symbols and the dashed line indicate their best-fit relation. We used the line dispersion of the \Ox\ line (green open square) in the unresolved NLR of HE~2158$-$0107 as a surrogate for $\sigma_*$, following the suggestion by \citet{Nelson:2000}.}
 \label{he2158_fig:bulge_relations}
\end{figure*}

Assuming that the velocity field is due to a dynamically cold rotating gas disc within the bulge of the host galaxy, we can infer a rough estimate of the dynamical mass $M_\mathrm{dyn}$ via
\begin{equation}
 M_\mathrm{dyn}=\frac{R\,v_r^2}{G\sin(i)}\ ,
\end{equation}
where $i$ is the inclination of the gas disc with respect to our line-of-sight, $R$ is the radius at which the radial velocity $v_r$ is measured, and $G$ is the gravitational constant. The maximum of the radial velocity is reached somewhere between 2\arcsec\ (7\,kpc) and 1\arcsec\ (3.5\,kpc), at which nearly all the mass of the bulge should be included considering its low effective radius of $r_\mathrm{e}\sim1.4\,\mathrm{kpc}$ in the $z$ band.  Since the host galaxy has no prominent stellar disc we use the axis ratio of the EELR $b/a=0.59$ to estimate an inclination of $i=55\degr$. We assign a large error of $\pm30\degr$ to this inclination estimate, because the unknown geometry of the AGN ionisation cones with respect to the presumed gas disc can introduce a large systematic uncertainty to the observed $b/a$ ratio \citep{Mulchaey:1996b}. With these very crude assumptions we estimate a possible range in $M_\mathrm{dyn}$ of $\sim$\,$ 3\times10^9$--\,$4\times10^{10}\mathrm{M}_{\sun}$, which is very well consistent with our SED-based upper limit on the stellar mass.

We also tried to estimate the velocity dispersion $\sigma_*$ of the bulge. Although we cannot directly determine $\sigma_*$ from the PMAS spectrum of the QSO, \citet{Nelson:1996} found that in Seyfert galaxies, the dispersion of the \Ox\ emission line in an unresolved NLR is correlated with $\sigma_*$.  Thus, \citet{Nelson:2000} suggested that $\sigma_{[\ion{O}{iii}]}$ may be used as a surrogate for $\sigma_*$. The \Ox\ line in the unresolved NLR of HE~2158$-$0107 can be described by a single Gaussian profile with a line dispersion of $\sigma_{[\ion{O}{iii}]}= 128\pm11\,\mathrm{km\,s}^{-1}$ after subtracting the instrumental resolution in quadrature.

\subsection{Is HE~2158$-$0107 an outlier from the BH-bulge relations?}
Above we collected several independent estimates for the bulge properties of the QSO host.  Figure~\ref{he2158_fig:bulge_relations} highlights the position of HE~2158$-$0107 in comparison to the $M_\mathrm{BH}$-$L_{\mathrm{bulge},H}$ relation of \citet{Marconi:2003} (left panel), the $M_\mathrm{BH}$-$M_{\mathrm{bulge}}$ relation of \citet{Haering:2004}, and the $M_\mathrm{BH}$-$\sigma_{*}$ relation studied by \citet{Gueltekin:2009}. We find that this QSO host is offset from the local scaling relations \emph{in all three cases}.

The bulge luminosity $L_{\mathrm{bulge,}H}$ is smaller by more than 0.6\,dex than the value indicated by the relation.  One advantage of the $H$ band is that the $K$-correction \citep[e.g.][]{Oke:1968,Hogg:2002} is almost negligible  \citep[e.g.][]{Mannucci:2001,Chilingarian:2010} up to $z\sim0.5$ and largely independent of the galaxy type. We therefore do not expect that the offset is caused by a systematic effect due to the redshift. Since we adopted a robust $3\sigma$ upper limit for $L_{\mathrm{bulge,}H},$ the true offset from the relation may even be substantially larger. 

It is clear that our estimate for $M_{\mathrm{bulge}}$ from the SED is tied to the upper limit in $L_{\mathrm{bulge,}H}$ and thus appears similarly offset from the $M_\mathrm{BH}$-$M_{\mathrm{bulge}}$ relation\footnote{Any stellar population model including contributions from TP-AGB stars \citep[e.g.][]{Maraston:2005} will result in lower stellar masses compared to the \citeauthor{Bruzual:2003} models used here. In this case the offset of HE~2158$-$0107 from the general mass scaling relation would even be increased, hence our use of \citet{Bruzual:2003} is conservative in this respect.} (middle panel of Fig.~\ref{he2158_fig:bulge_relations}). But also the estimated dynamical mass from the extended gas kinematics -- while uncertain -- is in agreement with the upper limit on stellar mass. Furthermore, the small apparent size of the bulge measured in the $z$ band supports the notion of an under-massive bulge with $M_{\mathrm{bulge}}=(1.4\pm0.6)\times10^{10}\mathrm{M}_{\sun}$ based on the empirical size-mass relation of bulge-dominated galaxies \citep{Shen:2003,Shen:2007}. It suggests that the bulge of HE~2158$-$0107 could be significantly away from $M_\mathrm{BH}$-$M_{\mathrm{bulge}}$ relation. Deeper high-spatial resolution photometry in the infrared and optical emission-line free bands are however required to accurately determine the SED of the host galaxy and to pin down the bulge luminosity and mass.

The offset in the BH mass-bulge relations appears to be  most conspicuous in $\sigma_*$. The measured value of $\sigma_{[\ion{O}{iii}]}$ is about half of the expected stellar velocity dispersion ($\sigma_*=245\pm44\,\mathrm{km\,s}^{-1}$) from the $M_\mathrm{BH}$-$\sigma_\mathrm{*}$ relation at the BH mass of HE~2158$-$0107 ($\log(M_\mathrm{BH}/\mathrm{M}_{\sun})=8.5\pm0.3$). Conversely, the measured velocity dispersion would correspond to a BH mass of $M_\mathrm{BH} = 2\times10^{7}\mathrm{M}_{\sun}$ adopting the $M_\mathrm{BH}$-$\sigma_*$ relation of \citet{Gueltekin:2009}, smaller than the expected value by more than an order of magnitude. The discrepancy remains high even when we account for a potential intrinsic offset $\Delta\sigma = \log(\sigma_*)-\log(\sigma_{[\ion{O}{iii}]})=0.11\,\mathrm{dex}$ as reported by \citet{Ho:2009a}. However, the robustness of $\sigma_{[\ion{O}{iii}]}$ as a surrogate for $\sigma_*$ is strongly debated as the scatter in the relation is much higher for $\sigma_{[\ion{O}{iii}]}$ than for $\sigma_*$ \citep{Shields:2003,Boroson:2005,Greene:2005b}.

Each measurement has some drawbacks and would certainly  be considered inconclusive individually. But remarkably, \emph{all} the independent results agree with each other and point into the same direction. Nevertheless, the offset of HE~2158$-$0107 from the $M_\mathrm{BH}$-bulge relations is still formally consistent with the intrinsic scatter of the scaling relations except for the velocity dispersion. So far we can refer to HE~2158$-$0107 only as a candidate outlier from the $M_\mathrm{BH}$-bulge relations. The fact that the QSO spectrum might be particularly suited to directly measure the \emph{stellar} velocity dispersion from a high S/N spectrum provides a good opportunity to verify and pin down the offset in the future. 

\subsection{The origin of the low-metallicity gas}\label{sect:tail}
\begin{figure}
 \resizebox{\hsize}{!}{\includegraphics[clip]{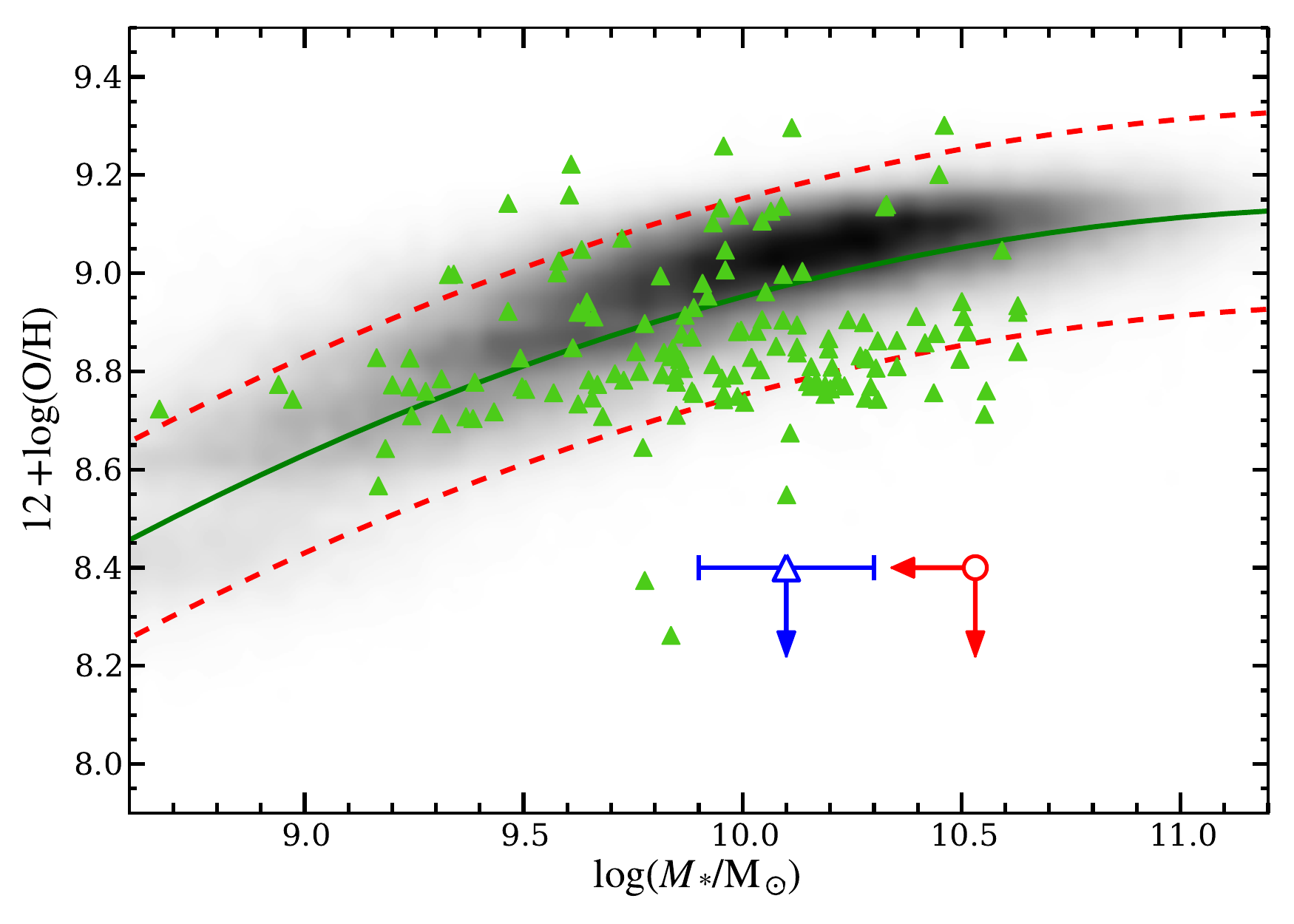}}
 \caption{Gas-phase oxygen abundance against total stellar mass.  The grey scale image correspond to a 2D histogram as constructed by \citet{Tremonti:2004} from the SDSS MPA/JHU galaxy catalogues. The red solid line represents the corresponding best-fit mass-metallicity relation for which the dashed lines indicate roughly its $3\sigma$ uncertainty. Most of the AGN hosts among the sample of \citet{Greene:2007} are within or slightly below this mass-metallicity relation. The blue open triangle and the red open circle highlight the exceptionally low metallicity of HE~2158$-$0107 with respect to the mass-metallicity relation, for which the stellar mass was estimated from the bulge size and the host SED, respectively.}
 \label{fig:mass_metal}
\end{figure}
While there is strong evidence for a significantly lower mass of the host galaxy than what would be expected from $M_\mathrm{BH}$, its mass is still too high to easily explain the low metallicity of the gas in the NLR and EELR. The mass-metallicity relation decreases only by 0.2\,dex from  $10^{11}\mathrm{M}_{\sun}$ to $10^{10}\mathrm{M}_{\sun}$ in stellar mass as shown in Fig.~\ref{fig:mass_metal}. Thus, the gas-phase NLR metallicity of HE~2158$-$0107 is still far below the mass-metallicity relation at the estimated stellar mass (limit) for this object. For comparison we also estimated stellar masses and gas-phase oxygen abundances for the AGN sample of \citet{Greene:2007}. We computed rough stellar masses from the absolute $g$ band host magnitudes with a mass-to-light ratio based on the $g-r$ colour \citep{Bell:2003}. Since only the total $g-r$ colour was reported by \citet{Greene:2007}, but not that of the host, we assumed $g-r\sim0.4$ as a typical value for blue-sequence galaxies \citep{Blanton:2003}. We estimated the gas-phase oxygen abundance with the calibration of \citet{Storchi-Bergmann:1998} only when the line ratios were above the \citet{Kewley:2001} demarcation line in the BPT diagram (cf. Fig.~\ref{he2158_fig:BPT_diagram}). For classical \ion{H}{ii} regions among the sample (i.e. objects below the \citet{Kauffmann:2003} line) we use the abundance calibration of \citet{Pettini:2004} and converted the measured oxygen abundance to the scale  of the \citeauthor{Tremonti:2004} calibration following the prescription of \citet{Kewley:2008}. Very few of the \citeauthor{Greene:2007} low BH mass AGN ($M_\mathrm{BH}<10^{6}M_{\sun}$) have oxygen abundances below $12+\log(\mathrm{O}/\mathrm{H})<8.5$ of which only three seem to have stellar masses comparable with that of HE~2158$-$0107 whereas all the others are consistent or only slight below the mass-metallicity relation. This highlights the exceptionally low gas-phase metallicity in HE~2158$-$0107 compared to other host galaxies of luminous AGN that have even much lower BH masses. We suspect that the large EELR with its intriguing tail-like geometry is the key to understand the unusual properties of HE~2158$-$0107.

\begin{figure}
 \resizebox{\hsize}{!}{\includegraphics[clip]{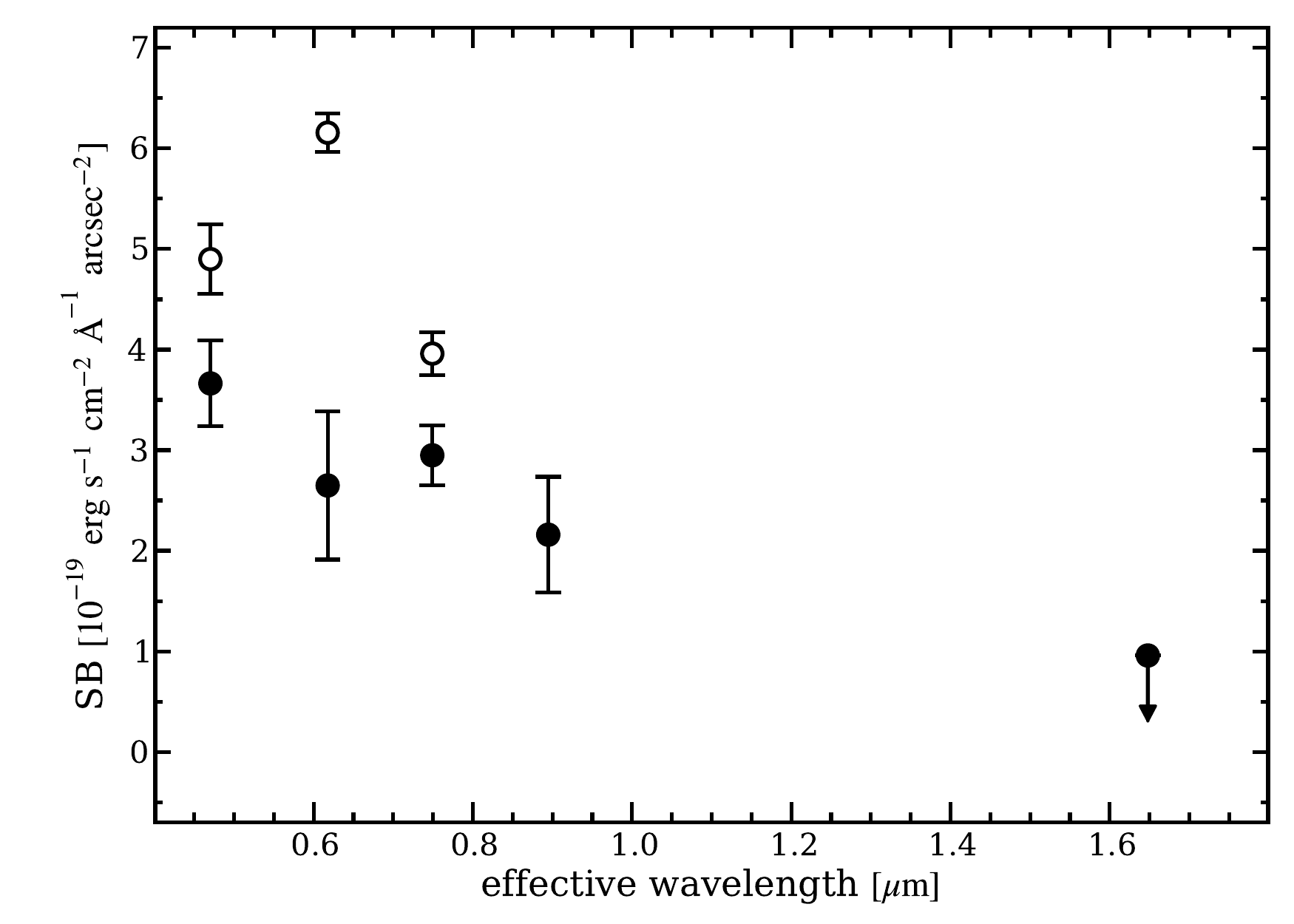}}
 \caption{SED of the extended tail close to HE~2158$-$0107. The surface brightness of a certain region centre on the extended tail (cf. Fig.~\ref{fig:host_models}) is shown at the effective wavelength of the corresponding passband. The initial surface brightnesses are indicated as the open symbols whereas the filled symbols denote the surface brightness after subtraction of the emission-line contribution. }
 \label{fig:SED_tail}
\end{figure}

The \Ox\ light distribution extends beyond the PMAS field of view, but we can follow the structure in the Stripe 82 $r$ and $i$ band images out to 30\,kpc projected distance from the host galaxy. We determined the emission-line contribution to the broad-band images from a small area at the South-East part of the EELR (see Fig.~\ref{fig:host_models}), which is still covered by the PMAS field of view but not  contaminated by host galaxy light. We constructed a synthetic emission-line spectrum for that area of the tail including the \Ox, \Hb\ and \Ha\ lines based on the observed \Ox\ flux, assuming an \Ox/\Hb\ line ratio of 7 and a Balmer decrement of 2.86. Furthermore,  we assumed an average [\ion{O}{ii}]/\Ox\ line ratio of the order of 1/3 \citep[e.g.][]{Villar-Martin:2008} to roughly predict the emission-line contamination in the $g$ band. The surface brightness values of the tail are reported in Table~\ref{tab:phot_tail} for both the SDSS photometry and the synthetic emission line spectrum. The corresponding SED is shown in Fig.~\ref{fig:SED_tail}. We find that the emission lines actually account for 57\% of the flux in the $r$ band whereas the \Ha\ line account for 25\% of the $i$ band flux. The emission lines are clearly dominating the $r$ band flux, but an underlying continuum must also be present with a corresponding \Ox\ rest-frame EW of $\mathrm{EW(\Ox)} = 1120\pm500$\,\AA. Such a high EW for the \Ox\ line was only observed in the integrated light of 2 out of 256 type 2 QSOs \citep{Zakamska:2003}. If the EELR should be powered by star formation, only blue compact dwarf galaxies reach such high equivalent widths in the \Ox\ line \citep[e.g.][]{Thuan:2005}.

\begin{table}
 \centering
 \caption{Photometry of the extended tail}
  \label{tab:phot_tail}
 \begin{tabular}{ccc}\hline\hline\noalign{\smallskip}
Band  &  SDSS image & Line Spectrum  \\ 
   &  [$\mathrm{mag\, arcsec}^{-2}$] & [$\mathrm{mag\, arcsec}^{-2}$]  \\ \noalign{\smallskip}\hline\noalign{\smallskip}
$g$ & $25.01\pm0.07$ & $26.50\pm0.20$ \\
$r$ & $24.17\pm0.03$ & $24.78\pm0.16$ \\
$i$ & $24.23\pm0.06$ & $25.71\pm0.20$ \\
$z$ & $24.50\pm0.26$ & (...) \\
$H$ & $>24.42$ & (...) \\
\noalign{\smallskip}\hline\end{tabular}

\end{table}

One explanation for this intriguing feature dominated by gaseous emission may be that it is a tidal tail due to a recent galaxy interaction. This scenario would be attractive for HE~2158$-$0107, because the oxygen abundance in interacting galaxies has been found to be systematically lower at a given stellar mass \citep{Ellison:2008, Michel-Dansac:2008,Alonso:2010}, presumably due to inflowing low-metallicity gas from larger radii. Among those, the strongest dilution of metallicities has been observed and theoretically predicted for major mergers  \citep{Rupke:2010,Montuori:2010}.
However, while this process can effectively funnel low-metallicity gas towards the galaxy nucleus, we find no signatures for a recent or ongoing major merger from the stellar light distribution at our spatial resolution of 1\,kpc/pixel. Pointing to a different stage of galaxy interactions, \citet{Villar-Martin:2010} recently reported the detection of a gas-rich tidal bridge of 180\,kpc length between a type 2 QSO ($z=0.399$) and a companion galaxy. The gas along that bridge is ionised due to ongoing star formation, but the QSO is capable of ionising the gas in its vicinity up to $\sim$35\,kpc distance. The EW of the \Ox\ emission and the associated low gas-phase metallicity along the tidal bridge of that interacting system are consistent with the EELR properties of HE~2158$-$0107. However, we cannot identify a potential companion galaxy around HE~2158$-$0107 that could be responsible for such a tail.

To put constraints on the mass of a potential interacting companion we estimated the ionised gas mass from the emission line luminosity of the entire EELR. Following the prescription given in \citet{Osterbrock:2006} and assuming Case B recombination, a low-density limit and a gas temperature of $T\approx 10,000$\,K for ionised nebulae, the ionised gas mass is given by
\begin{equation}
 M_{\mathrm{ion}} \approx \left(\frac{100\,\mathrm{cm}^{-3}}{n_\mathrm{e}}\right)\left(\frac{L_{\mathrm{H}\beta}}{10^{41}\,\mathrm{erg}\,\mathrm{s}^{-1}}\right) 10^{7}\mathrm{M}_{\sun}\ ,
\end{equation}
where $n_\mathrm{e}$ is the electron density of the gas. Since $n_\mathrm{e}$ cannot be estimated from our data, we assume a density of the order of $100\,\mathrm{cm}^{-3}$, which is a typical value for bright EELR around QSOs on kpc scales \citep[e.g.][]{Stockton:2002}. The \Ox\ luminosity of the entire EELR is $L_{[\ion{O}{iii}]}=2.1\times10^{42}\mathrm{erg\,s}^{-1}$, which corresponds to an \Hb\ luminosity of $L_{\mathrm{H}\beta}=3\times10^{41}\mathrm{erg\,s}^{-1}$ assuming an average \Ox/\Hb\ ratio of 7 throughout the entire EELR. Under these assumptions we estimate an ionised gas mass of roughly $M_{\mathrm{ion}}\approx 3\times10^{7}\mathrm{M}_{\sun}$. This is not a large amount of ionised gas even though cold molecular and neutral gas could increase the total gas mass by an order of magnitude. A tidally disrupted small gas-rich companion would thus be sufficient to account for the observed ionised gas mass.

Considering the exceptionally low metallicity of the gas in the NLR of this luminous QSO, the accretion of \emph{external} and \emph{nearly pristine} gas from the environment is the most plausible explanation to account for the strong dilution of the gas-phase oxygen abundance at the galaxy centre. Our limits on the oxygen abundance are actually in a regime where nitrogen may be expected to be predominantly of primary origin for which the used calibration of \citeauthor{Storchi-Bergmann:1998} is not valid any more, but would then even strengthen our suggestion that the inflowing gas is nearly pristine.  Whether the infall of gas is caused by interactions with small satellite galaxies or by the ``smooth'' accretion of gas is difficult to distinguish, because our observations do not provide enough spatial resolution to resolve substructures in the EELR and are not deep enough to reliable detect any related underlying stellar continuum.

\section{Discussion and Conclusions}
Combining PMAS integral field spectroscopy and multi-colour broad band images, we discovered several interesting properties of the luminous QSO HE~2158$-$0107. We find strong evidence for a sub-solar metallicity of its NLR. Only very few AGN are currently known to have such low NLR metallicities, and all of them are low mass system with much lower BH masses than the value found for HE~2158$-$0107, $\log(M_\mathrm{BH}/\mathrm{M}_{\sun})=8.5\pm0.3$. This QSO clearly is a highly unusual object at least at low redshift.

Another interesting property of this QSO is that its host galaxy appears to be offset from basically all known BH-bulge relations established for local inactive galaxies. The stellar mass and $H$ band luminosity, inferred from the SED of the bulge-dominated host, are both at least $0.6$\,dex lower than expected, which is independently supported by the kinematics of the ionised gas. Clearly, the offset for a single object can always be explained by chance given the intrinsic scatter of the relation, but the low stellar mass estimated from the size-mass relation of galaxies and the low velocity dispersion of the nuclear gas suggest that HE~2158$-$0107 is an outlier from the local $M_\mathrm{BH}$-bulge relations. We tentatively note that the estimated offset from the relation matches, by order of magnitude, with the systematic offsets found at higher redshifts \citep[e.g.][]{Peng:2006,Treu:2007,Schramm:2008,Merloni:2010}. \citet{Jahnke:2010} recently studied with numerical simulations how the BH mass-bulge relations may get established over cosmic time. They found that merging of galaxies in a $\Lambda$CDM universe alone is sufficient to produce a tight relation at $z=0$, even if the BH and bulge masses are initially uncorrelated. In fact, the local relation would have a much lower scatter than actually  observed if no additional physical processes operate that \emph{increase} the scatter. Thus, it is important to understand the physical mechanisms that lead to the observed scatter in the BH mass-bulge relations. It may be that HE~2158$-$0107 is currently offset from the relation due to such a process or event that influences the relative growth of BH and bulges.

Whether or not the low-metallicity gas in this object is causally connected to the higher-than-expected $M_\mathrm{BH}$/$M_\mathrm{bulge}$ ratio is a key question. A striking feature of HE~2158$-$0107 is its large EELR with a tail-like geometry extending out to 30\,kpc from the QSO.  EELRs are common around radio-loud QSOs and are well-studied. However, it is still controversial whether they are produced by tidal debris from interacting galaxies, inflated by  galactic superwinds, or whether they may also represent smoothly accreted cold gas from the environment  \citep[e.g.][]{Stockton:1987,Fabian:1987,Keres:2005,Stockton:2006,Fu:2009}. In the case of our radio-quiet QSO, a galactic superwind can be excluded with high confidence, because the EELR kinematics are rather quiescent and a large scale radio jet is absent. A major merger origin of the extended tail is also unlikely considering its low ionised gas mass and the lack of obvious major merger signatures in the continuum. Instead we propose that the accretion of external low-metallicity gas explains the exceptional properties of this QSO and its host galaxy. This process supplies fuel for the growth of the bulge and presumably also of the BH.

The details of the presumed accretion process remain difficult to assess with our observations. One option is that the material is accreted from one, possibly several low-metallicity dwarf galaxies in a sequence of minor merger events. If so, this process must have been going on sufficiently long for substantial amounts of low-metallicity gas to reach the NLR of the QSO. An alternative, but also more speculative possibility is the smooth accretion of gas directly from the intergalactic medium, which would naturally explain the observed low metallicity of the gas. A fraction of the accreted gas needs to be dense enough in order to overcome the virial shock while falling onto the galaxy and to remain at relatively cool temperatures $\sim$\,$10^4$\,K \citep[e.g.][]{Birnboim:2003,Keres:2005}, which would then light up as an EELR. This process is theoretically predicted, but observational evidence for its existence is still rare with only a few candidate detections  \citep{Giavalisco:2011,Ribaudo:2011}, also because the phenomenon is expected to be more common at high redshift than in the local universe. To shed more light on this issue, much deeper spectra covering the entire optical window are needed to turn our currents limits into robust abundance measurements that can then be directly compared with those of dwarf galaxies. In this respect a study of the stellar metallicity of the host galaxy via absorption lines appear much more difficult given its compactness and the blending with the QSO. On the other hand, the QSO provides a bright background source to measure abundance of different elements like O, N and C directly from rest-frame ultraviolet metal absorption lines of the gas that could be compared with the abundance pattern for dwarf galaxies or expectations for the inter galactic medium at this low redshift.

Independently of the question whether the accretion process is ``smooth'' or ``clumpy'', in our preferred interpretation HE~2158$-$0107 represents a particular phase of substantial galaxy and BH growth that can be observationally linked with the accretion of external gas. A fundamental difference in the relative BH and bulge growth might potentially occur when a BH is  stochastically fed by an intrinsic gas reservoir within its host galaxy in contrast to this process where large amounts of nearly pristine gas are dumped onto the galaxy. More detailed observations are certainly needed to further constrain the properties of this intriguing object and to elucidate on its importance as a ``laboratory'' for studying such a particular phase in the evolution of massive galaxies.

\begin{acknowledgements}
We thank the anonymous referee for valuable comments and suggestions that improved the quality of the present paper.
We are grateful to Jakob Walcher for sharing his knowledge and experience with stellar population modelling.
BH and LW acknowledge financial support by the DFG Priority Program 1177 ``Witnesses of Cosmic History: Formation and evolution of black holes, galaxies and their environment'', grant Wi 1369/22.
KJ acknowledges support through the DFG Emmy Noether-Program, grant JA 1114/3-1.

Funding for the SDSS and SDSS-II was provided by the Alfred P. Sloan Foundation, the Participating Institutions, the National Science Foundation, the U.S. Department of Energy, the National Aeronautics and Space Administration, the Japanese Monbukagakusho, the Max Planck Society, and the Higher Education Funding Council for England. The SDSS was managed by the Astrophysical Research Consortium for the Participating Institutions.

This research has made use of the NASA/IPAC Extragalactic Database (NED) which is operated by the Jet Propulsion Laboratory, California Institute of Technology, under contract with the National Aeronautics and Space Administration. 

For the preparation of this paper we have made used of the cosmology calculator ``CosmoCalc'' \citep{Wright:2006}.
\end{acknowledgements}
\bibliographystyle{bibtex/aa}
\bibliography{references}
\appendix
\section{Estimation of the \Ni\ detection limit in the spectra}
The \Ni\ $\lambda\lambda 6548,6583$ doublet could neither be detected in the QSO nor in the EELR spectra obtained for HE~2158$-$0107 with the PMAS spectrograph. Due to the diagnostic importance of this line we determined robust upper limits for those lines with the aid of dedicated Monte Carlo simulations. 

\begin{figure}
\resizebox{\hsize}{!}{\includegraphics[]{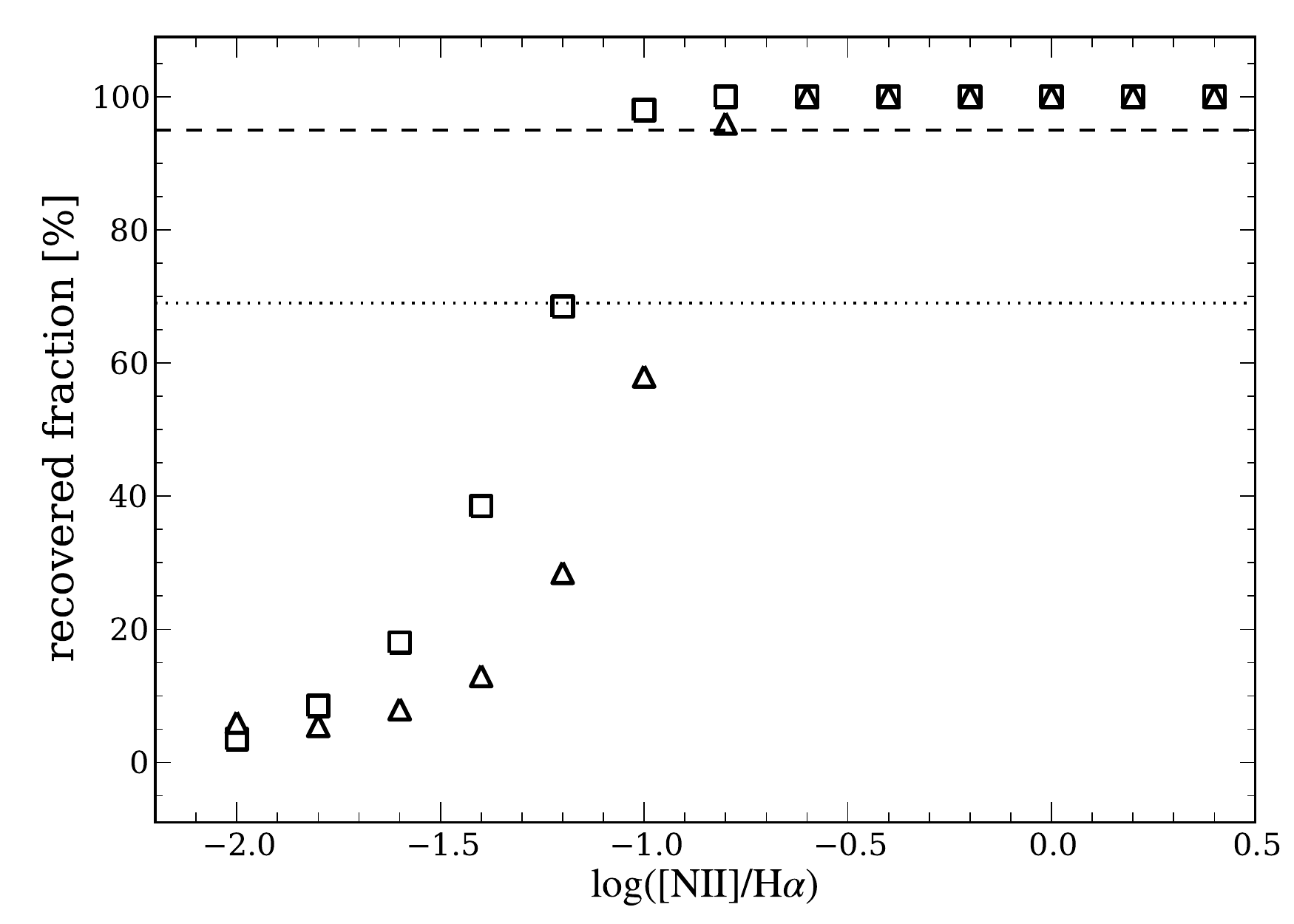}}
\caption{Fraction of $3\sigma$ \Ni\ detections among 200 simulated mock spectra as a function of simulated \Ha/\Ni\ ratio. The open squares correspond to simulated QSO spectra and the open triangles correspond to simulated EELR spectra, respectively. The dashed and solid lines highlight the 95\% and 68\% confidence limits corresponding to a $3\sigma$ and $1\sigma$ detection limit, respectively.}
\label{he2158_fig:N2_sim}
\end{figure}

Based on the best-fit \Ha\ model for the QSO and EELR spectrum we generated mock spectra including narrow \Ni\ $\lambda\lambda 6548,6583$ lines, with the same profile and redshift as the narrow \Ha\ line, within a certain range in \Ha/\Ni$\lambda 6583$ ratios (\Ha/\Ni\ hereafter) . We further added realistic Gaussian noise to the spectra including night sky lines using the variance vector matched to the observed data. 200 mock spectra were generated for each \Ha/\Ni\ ratio.

In order to test for each mock spectrum whether the \Ni\ lines could be recovered at a certain confidence level, we fitted the spectra with two different models, with and without the \Ni\ components. The \Ni\ lines actually add only one additional free parameter to the model, the \Ni$\lambda 6583$ flux, as the \Ni\,$\lambda6548$/\Ni\,$\lambda6583$ flux ratio is fixed to 1/3 and the line width and redshift were both coupled to the narrow \Ha\ component. We employed the statistical F-test to check whether the model including the \Ni\ is a significantly better model than the one without based on the $\chi^2$ values for a given simulated spectrum at a 95\% confidence level. 

In Fig.~\ref{he2158_fig:N2_sim} we present the fraction of spectra for which the model including a \Ni\ component was significantly better representing the data as a function of the input \Ha/\Ni\ line ratio. At high \Ha/\Ni\ ratios, the \Ni\ is detected in 100\% of the spectra. This fraction drops at a certain \Ha/\Ni\ ratio below 95\% or 69\%, which we adopt as our $3\sigma$ and $\sigma$ detection limits, and smoothly decreasing towards zero with decreasing \Ha/\Ni\ ratio as expected. Our simulations imply a $3\sigma$ upper limit of $\Ha/\Ni<-0.8$ for the EELR spectrum ($\Ha/\Ni<-0.8$ 1$\sigma$) and $\Ha/\Ni<-1.0$ for the QSO spectrum ($\Ha/\Ni<-1.2$ 1$\sigma$), respectively.

\section{Systematics of the QSO-host decomposition}
\begin{figure*}
 \includegraphics[width=\textwidth,clip]{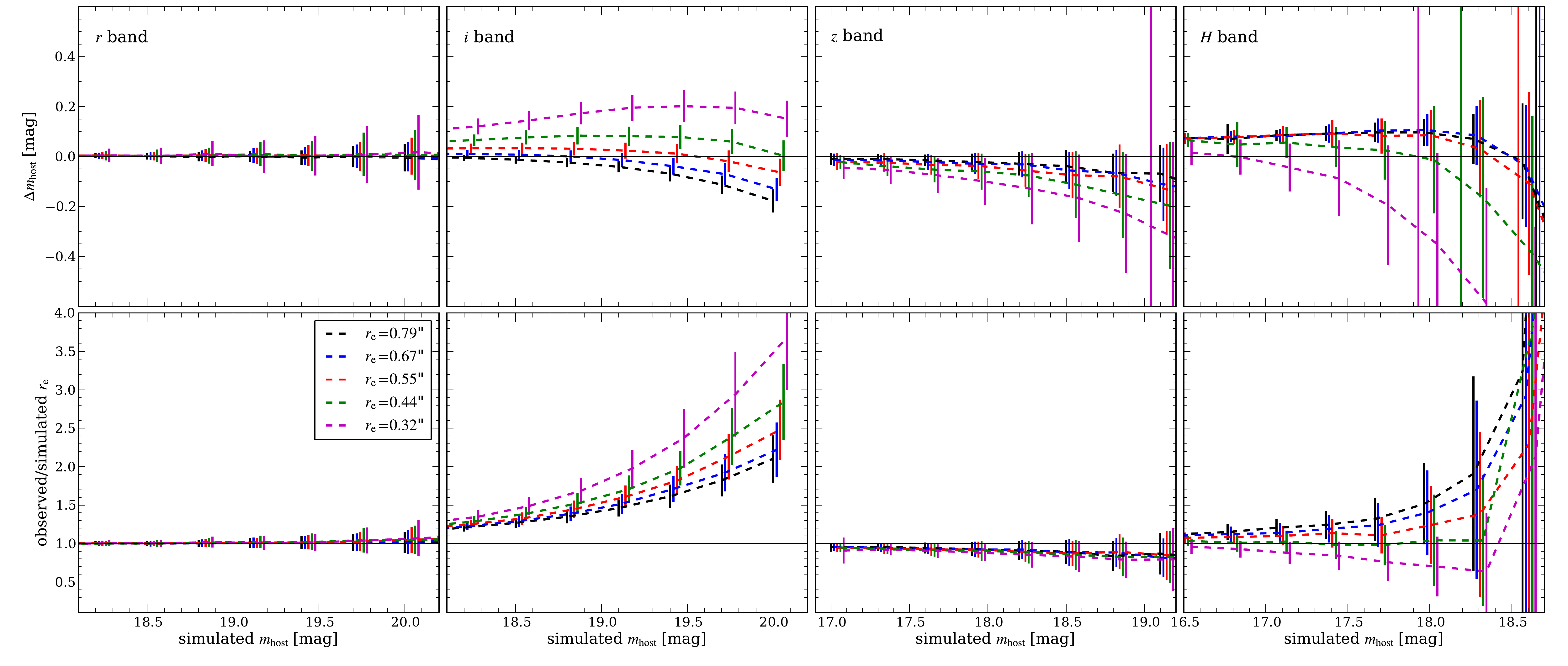}
 \caption{Results of Monte Carlo simulations to estimate the systematic uncertainties of the 2D decomposition with \texttt{GALFIT} for HE~2158$-$0107. The upper panels present the difference in simulated and observed host magnitudes  ($\Delta m_\mathrm{host}$) as a function of the simulated host magnitude for 5 different effective radii assuming a de Vaucouleurs surface brightness profile. Detection limits (3$\sigma$) in host brightness for the different effective radii are indicated by the solid vertical lines in their corresponding colour. The lower panels show the ratio of the observed over the simulated effective radii as a function of the simulated host brightness.}
  \label{he2158_fig:host_simulations}
 \end{figure*}
The two-dimensional QSO-host decomposition is a complex process which is not free of systematic effects, especially in the high contrast regime. Here we present Monte Carlo simulation designed to investigate these systematic effects specifically for HE~2158$-$0107. We infer realistic errors for each band and obtain a robust upper limit for the host brightness in the $H$ band.
We created a suite of mock images for each band with a fixed QSO magnitude corresponding to the observation. The surface brightness (SB) distribution of the underlying host galaxy was set to a de Vaucouleur law, the effective radii $r_\mathrm{e}$ were set to $0\farcs32$, $0\farcs44$, $0\farcs55$, $0\farcs67$ or $0\farcs79$, and the host magnitudes varied within a reasonable range in steps of 0.3\,mag. We produced 200 mock images for each set of parameters from the two-dimensional grid in parameter space. The  host orientations and the axis ratios ($b/a$) were randomly chosen where $b/a$ was restricted to a range of 0.5--1.0. The simulated images were convolved with one of the available two PSF stars and the Gaussian noise pattern of the observed images was added before the images were analysed with \texttt{GALFIT} using the other PSF star. 

The results of the Monte-Carlo simulations, shown in Fig.~\ref{he2158_fig:host_simulations}, reveal that each band is subject to different systematic effects. Obviously, the $r$ band is the deepest image of the four with the highest S/N introducing no discernable systematics. For the $i$ band image, we found that $r_\mathrm{e}$ increases strongly with the nucleus-to-host ratio, so that the $r_\mathrm{e}$ was significantly overestimated.  We suspect that this could be caused by a PSF mismatch between the two PSFs for this band. The recovered $r_\mathrm{e}$ in the $z$ and $H$ bands displays only little variation with host magnitude, but appears to be slightly offset from the true input value.

The uncorrected host magnitudes have substantial systematic errors that can only be determined with these simulations. Again the $i$ band exhibits the strongest deviations from the input values. In all cases, except the $r$ band, we found that $\Delta m_\mathrm{host}$ is decreasing with increasing host brightness, which corresponds to a flux transfer from the PSF to the host component. The systematic offsets and errors matched to the observed parameters of the host in the $r$, $i$ and $z$ bands are $\Delta m_{\mathrm{host},r}=0.00\pm0.02$\,mag, $\Delta m_{\mathrm{host},i}=0.08\pm0.04$\,mag and $\Delta m_{\mathrm{host},z}=-0.17\pm0.11$, respectively.  The recovered host fluxes were in all bands significantly above the 3$\sigma$ detection limit, except in the $H$ band. The $3\sigma$ detection limit for the host in the this band depends critically on $r_\mathrm{e}$ and corresponds to $m_{\mathrm{host},H} >18.19$ for $r_\mathrm{e}=0\farcs44$.

\end{document}